\documentclass[conference]{IEEEtran}
\IEEEoverridecommandlockouts
\usepackage{cite}
\usepackage{amsmath,amssymb,amsfonts}
\usepackage{algorithm,algpseudocode}
\usepackage{graphicx}
\usepackage{textcomp}
\usepackage{xcolor}
\usepackage{subcaption}
\usepackage{multirow}
\usepackage{capt-of}
\usepackage[draft]{pgf}
\definecolor{azure}{rgb}{0.0, 0.5, 1.0}

\def\BibTeX{{\rm B\kern-.05em{\sc i\kern-.025em b}\kern-.08em
    T\kern-.1667em\lower.7ex\hbox{E}\kern-.125emX}}
\begin{document}

\title{ViPer NL-COMM: Making Vector Perturbation Precoding Practical

}

\author{\IEEEauthorblockN{Thomas James Thomas, George N. Katsaros, Chathura Jayawardena and Konstantinos Nikitopoulos}
\IEEEauthorblockA{{6G Innovation Centre, Institute for Communication Systems} \\
{University of Surrey}
Guildford GU2 7XH, Surrey, United Kingdom \\}
}
\maketitle

\begin{abstract}
Large multiple-input multiple-output (MIMO) systems rely on efficient downlink precoding to enhance data rates and improve connectivity through spatial multiplexing. However, currently employed linear precoding techniques, such as minimum mean square error (MMSE) precoding, significantly limit the achievable spectral efficiency. To meet practical error-rate targets, existing linear methods require an excessively high number of access point (AP) antennas relative to the number of supported users, leading to disproportionate increases in power consumption. 
%
Efficient non-linear processing frameworks for uplink MIMO transmissions, such as NL-COMM, have been proposed. However, downlink non-linear precoding methods, such as Vector Perturbation (VP), remain impractical for real-world deployment due to their exponentially increasing computational complexity with the number of supported MIMO streams.
This work presents ViPer NL-COMM, the first practical algorithmic and implementation framework for VP-based downlink precoding. ViPer NL-COMM extends the core principles of NL-COMM to the precoding problem, enabling scalable parallelization and real-time computational performance while maintaining the substantial spectral-efficiency benefits of VP precoding.
ViPer NL-COMM consists of a novel mathematical framework and an FPGA prototype
capable of supporting large MIMO configurations (up to 16×16), high-order modulation (256-QAM), and wide bandwidths (100 MHz) within practical power and resource budgets.
System-level evaluations demonstrate that ViPer NL-COMM achieves target error rates using only half the number of transmit antennas required by linear precoding, yielding net power savings on the order of hundreds of Watts at the RF front end. Moreover, ViPer NL-COMM enables supporting more information streams than available AP antennas when the streams are of low-rate, paving the way for enhanced massive-connectivity scenarios in next-generation wireless networks.

\end{abstract}


\section{Introduction}
Future-generation wireless networks promise ubiquitous connectivity with peak data rates tens to hundreds of times greater than current standards \cite{giordani'20,you'23}. Meeting these ambitious goals requires significant enhancements in spectral efficiency, beyond merely expanding the transmission bandwidth\cite{nik'22}. 
In this context, multiple-input multiple-output (MIMO) technology \cite{larss'14} plays a central role, enabling the concurrent transmission of multiple data streams over the same time-frequency resources through spatial multiplexing. By exploiting spatial degrees of freedom, MIMO systems have become essential to meeting the growing capacity and throughput demands of emerging wireless applications \cite{saad'20}.

Conventional MIMO deployments typically employ linear processing techniques, which transform the MIMO channel into several independent parallel channels. These techniques are widely favored because of their low computational complexity and seamless integration with existing hardware. However, linear precoding methods such as Zero-Forcing (ZF) \cite{wiesel'08} and Minimum Mean Square Error (MMSE) \cite{hochwaldi'05} lead to significant underutilization of channel capacity \cite{castan'13}. As a result, and as we show later in detail, supporting even a small number of concurrently transmitted streams requires a disproportionately large number of antennas and radio frequency (RF) chains at the BS, even when the streams are of low rate. This inefficiency directly impacts energy consumption and limits the ability of the system to support dense user deployments within practical power budgets.

Existing Non-linear (NL) precoding techniques can substantially outperform linear methods, but each faces unique practical challenges. For instance, Dirty Paper Coding (DPC) \cite{costa'83} theoretically achieves full channel capacity but remains computationally prohibitive. Tomlinson-Harashima Precoding (THP) \cite{harashima'72} reduces complexity by iteratively canceling inter-user interference; however, it still underutilizes available channel capacity and suffers from propagation errors in multi-user scenarios. Vector Perturbation (VP), an extension of THP, improves upon this by optimally perturbing the transmitted signal constellation, maximizing the receiver's signal-to-noise ratio (SNR). However, VP involves solving an NP-hard lattice optimization problem, significantly complicating real-time implementation.
Sphere encoding (SE) techniques, particularly those employing highly optimized depth-first strategies \cite{nik'14,SD_statitical,SD_VLSI}, reduce computational load by efficiently pruning the search space. Yet, their inherently sequential structure significantly limits parallelism, constraining computational throughput. Fixed-complexity SE methods, such as K-Best and Fixed-Complexity Sphere Encoders (FCSE) \cite{mohaisen'11,habendorf2007vector}, introduce controlled parallelization to achieve consistent throughput. 
K-Best utilizes breadth-first search, balancing complexity and performance.
Nonetheless, achieving practical error-rate performance often necessitates increasing the value of the search parameter 
$K$, which in turn raises the sorting overhead and severely impacts computational throughput.
On the other hand, FCSE avoids sorting altogether by fixing the number of explored nodes per layer, thereby achieving predictable latency suitable for hardware implementation. However, this comes at the cost of adaptability to dynamic channel conditions and varying processing requirements.

While non-linear precoding in the downlink remains hindered by the above-mentioned challenges, significant advances have been made in the context of uplink non-linear MIMO detection, where similar optimization challenges arise. 
The Non-Linear Processing for Communications (NL-COMM) framework \cite{nlcommdotcom,nikitopoulos2019parallel}  has proposed practical solutions for low-latency and highly parallelizable MU-MIMO detection \cite{nik'18,nik'22,katsaros'24,nikaccess'21,JayawardenaChathura2024NDGo,husmann_flexcore_2017}.
Its key principles involve identifying sphere decoder tree paths deemed ``\emph{most promising}" for including the correct solution before the actual data transmission \cite{nik'18}. Building upon this, and combining an approximate version of the path selection process with parallel evaluation of candidate paths \cite{husmann_flexcore_2017},  NL-COMM achieves latencies comparable to linear detection methods while maintaining fixed throughput, even when implemented on commercial off-the-shelf (COTS) CPUs~\cite{katsaros'24}.

In this work, we introduce \textbf{V}ector Perturbation for H\textbf{i}ghly \textbf{Per}forming Precoding \textit{(ViPer) NL-COMM},\footnote{The initial results of ViPer were introduced in our earlier conference publication~\cite{husmann'18}.} the first practical MU-MIMO downlink design that efficiently implements VP in a flexible and massively parallel manner, building on some of the key principles from previous approaches in the uplink \cite{nik'18,husmann_flexcore_2017}. ViPer NL-COMM incorporates several innovative features to address existing challenges in VP precoding:
\begin{itemize}
    \item A preprocessing step to identify the most promising perturbation vectors before the data transmission, that maximize the received SNR. 
    For this purpose, we define a new Metric of Promise appropriate for the downlink. The preprocessing step relies solely on channel state information, allowing ViPer NL-COMM to focus the available processing power on the most promising perturbation vectors and thereby reduce the complexity of data vector-based post-processing.
    \item 
    Support for low-rate overloaded scenarios with more simultaneously transmitted data streams than BS antennas, enabling massive connectivity use cases. 
    \item A novel sorted RQ decomposition that significantly simplifies the preprocessing by avoiding computationally intensive pseudo-inverse calculations, essentially halving the number of required complex multiplications.
\end{itemize}
ViPer NL-COMM demonstrates significant error-rate performance improvements over the existing precoding methods. Specifically, as detailed in Section II C, ViPer NL-COMM achieves approximately 100\% higher throughput compared to MMSE precoding \cite{hochwaldi'05} and 30\% higher throughput than existing non-linear precoders \cite{habendorf2007vector,mohaisen'11} of the same complexity. Moreover, as shown in later sections, ViPer NL-COMM can support up to twice as many concurrently transmitting users as BS antennas, enabling efficient operation in overloaded downlink scenarios.

Beyond the novelties in the NL-COMM mathematical framework, this work also, for the first time, presents a complete hardware-efficient VP architecture and an FPGA implementation, fully integrating both preprocessing and perturbation vector computation stages. This enables, for the first time, a realistic evaluation of the framework’s computational performance and power consumption under practical hardware constraints.
Our FPGA-based design employs interleaved pipelining and spatial parallelization to substantially reduce processing latency, while supporting configurable parallelism to balance throughput and resource utilization as the MIMO system size scales.
Implemented on a Xilinx Virtex UltraScale+ XCVU9P FPGA, the proposed ViPer NL-COMM accelerator meets the real-time latency targets supporting large-scale MIMO configurations (up to 16×16), high modulation orders (256-QAM), and wide bandwidths (up to 100 MHz).
Finally, this study is the first to quantify system-level power efficiency gains of ViPer NL-COMM. Our results show that with an FPGA power overhead in the order of 10–20 Watts, ViPer NL-COMM can achieve nearly 200 Watts of power savings per base station compared to traditional linear methods, by significantly reducing the required RF elements and without performance loss compared to MMSE. As we later discuss in detail, these gains can be directly leveraged in systems with dynamic RF channel reconfiguration capabilities without modifications in the employed antenna array, addressing critical flexibility and energy-efficiency needs for future intelligent wireless networks.
Overall, ViPer NL-COMM represents a significant step toward practical, flexible, and power-efficient non-linear precoding for next-generation wireless infrastructure.

The remainder of this paper is organized as follows. Section II discusses the related work to ViPer NL-COMM in more detail. Section III introduces the system model for traditional precoding approaches and the ViPer NL-COMM approach. Section IV presents the architecture design for the ViPer NL-COMM algorithm with a high-level discussion of the key building blocks and design considerations. Section V evaluates the performance of the ViPer NL-COMM based FPGA implementation in comparison to existing designs, and the work is concluded in Section VI.

\section{System Model}
Consider an information vector $\mathbf{v}$ comprising quadrature amplitude modulation (QAM)-based symbols that undergo precoding to produce the transmit vector, denoted as $\mathbf{x}$. Assuming a flat fading MIMO channel realized by $N_T$ AP antennas and $N_R$ single-antenna users. The received vector can be mathematically represented as 
\begin{equation}
	\mathbf{y}=\mathbf{Hx} +\mathbf{n}
\end{equation}
where $\mathbf{H}$ is the downlink channel matrix with each of its entries $H_{i,j}$ denoting the channel coefficients between the $i^\textnormal{th}$ user and the $j^\textnormal{th}$ AP antenna. Here, $\mathbf{n}$ represents the additive Gaussian noise with each entry belonging to the normal distribution $\mathcal{CN}(0,\sigma^2)$. Owing to the finite transmit power $P_T$ of the wireless access points (APs), the precoded vector $\mathbf{x}$ needs to be constrained as $\mathbb{E}\{\mathbf{x}^H \mathbf{x}\}=P_T$.

While this work considers single-antenna user terminals, the proposed ViPer NL-COMM framework can be extended to multi-antenna users. In such cases, multiple data streams per user can be supported by applying lightweight intra-UE linear precoding (e.g., SVD-based beamforming) to decouple the UE’s spatial dimensions, after which ViPer operates across the aggregated streams.  In this direction, multi-antenna combining techniques \cite{dao2010optimizing,park2011mmse} can be applied and complement ViPer-NLCOMM when multi-antenna UEs are considered. However, determining the optimal strategy, optimizing the performance complexity tradeoff, and the detailed design and evaluation of such multi-antenna UE extension is beyond the scope of this work.

\subsection{Linear Precoding}
Linear (fully digital) downlink precoding transforms the information vector $\mathbf{v}$ by multiplying it with a precoding matrix, denoted as $\mathbf{G}$. As the objective of MMSE precoding is to maximize the signal-to-interference-and-noise-ratio (SINR) at the user terminals, the precoding matrix $\mathbf{G}_\textnormal{MMSE}$ is obtained by computing the regularized channel inverse as follows
\begin{equation}
\mathbf{G}_\textnormal{MMSE}=\mathbf{H}^H(\mathbf{H}\mathbf{H}^H+\alpha \mathbf{I})^{-1}
\end{equation}
where $\alpha$ is the regularization parameter and $\mathbf{I}$ is the identity matrix. To ensure that maximum transmit power is not exceeded, a normalization is performed on $\mathbf{x}$ by employing $\gamma_\textnormal{MMSE}$ such that $\gamma_\textnormal{MMSE}=\mathbb{E}\{\mathbf{x}^H\mathbf{x}\}=\mathbb{E}\{\mathbf{v}^H\mathbf{G}_\textnormal{MMSE}^H \mathbf{G}_\textnormal{MMSE} \mathbf{v}\}$. The received vector can now be represented as
\begin{equation}
\mathbf{y}_\textnormal{MMSE} = \mathbf{H} \sqrt{\dfrac{P_T}{\gamma_\textnormal{MMSE}}} \mathbf{G}_\textnormal{MMSE} \mathbf{v} + \mathbf{n}
\label{eq:gamm}
\end{equation}
\cite{hochwaldi'05} used eigenvalue decomposition to determine the conditional expectation of $\gamma_\textnormal{MMSE}$ with respect to $\mathbf{v}$ as\begin{equation}
\mathcal{E}_{\gamma_\textnormal{MMSE}} = \sum_{l=1}^{N_R} \dfrac{\lambda_l}{(\lambda_l+\alpha)^2}
\end{equation}
where $\lambda_l$ is the $l^\textnormal{th}$ eigenvalue of $\mathbf{HH}^H$. As \cite{hochwaldi'05} demonstrates, setting $\alpha=\sigma^2 N_R$ achieves maximum SINR and an optimal trade-off between very low signal powers and high multi-user interference. In practical systems the noise-plus-interference power of each user can be approximated from uplink measurements (e.g., reference-signal based) and/or from reported CQI.

\subsection{Generalized Vector Perturbation}
The goal of VP is to calculate a joint perturbation vector that maximizes the receivers' signal-to-noise-ratio (SNR). Considering $\gamma_\textnormal{VP}$ is the transmit power normalization factor in VP, the objective then becomes to minimize $\gamma_\textnormal{VP}$ by introducing a perturbation $\mathbf{t}$ \cite{hochwaldi'05}, consisting of complex entries of the form $(c+i\cdot d)$ where $c$ and $d$ are integers, to the information vector $\mathbf{v}$ and applying the pseudo-inverse of the channel matrix, given by $\mathbf{H}^{\dagger}=\mathbf{H}^H(\mathbf{HH}^H)^{-1}$. Let $\mathcal{T}$ denote the set of potential perturbation vectors. Considering the pre-normalization precoding vector $\mathbf{x}=\mathbf{H}^{\dagger}(\mathbf{v}-\tau\mathbf{t})$, where $\tau$ is the perturbation interval that is fixed for a given QAM order \cite{yuen'06}. The perturbation vector that maximizes SINR is given by the solution to the following optimization problem
\begin{equation}
{\tilde{\mathbf{t}}}=\arg\min_{\mathbf{t}\in \mathcal{T}}(\|\mathbf{H}^\dagger (\mathbf{v}-\tau \mathbf{t})\|^2)
\label{eq:se}
\end{equation}
Given that the precoded $\mathbf{x}$ needs to be scaled by $\sqrt{\tfrac{P_T}{\gamma_\textnormal{VP}}}$, determining $\gamma_\textnormal{VP}$ is not straightforward as it does not have a closed form expression, unlike $\gamma_\textnormal{MMSE}$. \cite{ryan'09} provides a lower bound estimate on $\gamma_\textnormal{VP}$, which was shown in \cite{husmann'18} to also be accurate for practical rate adaptation. The VP MIMO received vector becomes
\begin{align}
	\mathbf{y}_\textnormal{VP}&=\mathbf{H}\sqrt{\dfrac{P_T}{\gamma_\textnormal{VP}}} \mathbf{H}^\dagger (\mathbf{v}-\tau\mathbf{t})+\mathbf{n}\\
	&=\sqrt{\dfrac{P_T}{\gamma_\textnormal{VP}}}(\mathbf{v}-\tau\mathbf{t})+\mathbf{n}	
\end{align}
In particular, the normalization factor is given by 
$\gamma_{\mathrm{VP}} \;=\; \big\|\mathbf{H}^\dagger(\mathbf{v}-\tau\mathbf{t})\big\|^2$,
and the SINR of the received signal $\mathbf{y}_\textnormal{VP}$ is inversely proportional to $\gamma_\textnormal{VP}$, specifically expressed as $\frac{P_T}{\gamma_\mathrm{VP}\sigma^2}$. Therefore, minimizing $\gamma_\textnormal{VP}$ in Eq.~\ref{eq:se} effectively maximizes the SINR. The receiver can then simply employ the following modulo function
\begin{equation}
\tilde{\mathbf{y}}_\textnormal{VP}=\mod\left[\sqrt{\dfrac{\gamma_\textnormal{VP}}{P_T}}\mathbf{y}_\textnormal{VP}\right]_\tau
\label{eq:postproc}
\end{equation}
where $\mod\lbrack a \rbrack_b$ is the modulo-$b$ output with respect to $a$.

To optimally solve Eq. \ref{eq:se}, traditional SEs employ a QRD of the precoding matrix. Let $\mathbf{H}^\dagger=\mathbf{Q}\mathbf{R}$ with unitary $\mathbf{Q}$.
By unitary invariance of the Euclidean norm,
$\|\mathbf{H}^\dagger(\mathbf{v}-\tau\mathbf{t})\|^2
=\|\mathbf{Q}\mathbf{R}(\mathbf{v}-\tau\mathbf{t})\|^2
=\|\mathbf{R}(\mathbf{v}-\tau\mathbf{t})\|^2$,
to translate the $N_R$-dimensional integer-lattice least-squares minimization into a computationally less-complex tree search as shown below.
\begin{equation}
\label{eq_qrse}
\tilde{\mathbf{t}} \;=\; \arg\min_{\mathbf{t}\in\mathcal{T}}
\big\|\mathbf{R}(\mathbf{v}-\tau \mathbf{t})\big\|^2.
\end{equation}
Representing the partial perturbation vector of a node in the SE tree at level $k$ by $\mathbf{t}_k$, its partial Euclidean distance (PED) can be computed as
\begin{equation}
d(\mathbf{t}_k) \;=\; 
\left| \tilde{s}_k - \tau \sum_{l=k}^{N_R} R_{k,l}\, t_l \right|^2 
\;+\; d(\mathbf{t}_{k+1}), \quad k=N_R,\ldots,1,
\end{equation}

where $R_{k,l}$ denotes the $(k,\!l)$ entry of $\mathbf{R}$ (row $k$, column $l$).
The vector $\tilde{\mathbf{s}}$ is obtained by multiplying $\mathbf{R}$ with $\mathbf{v}$ and $d(\mathbf{t}_{N_R+1})$ is equal to zero. This simplifies the integer-lattice least-squares problem into finding the node with the smallest $d(\mathbf{t}_1)$.

\section{ViPer NL-COMM Design \& Evaluation}
This section discusses the key aspects of the ViPer NL-COMM digital precoder.
\subsection{ViPer NL-COMM design}

\subsubsection{Improved sorted RQ decomposition}
Traditional SEs translate (\ref{eq:se}) into a tree search problem by QR decomposing the precoding matrix $\mathbf{H}^\dagger$ as $\mathbf{H}^\dagger=\mathbf{QR}$. It is well known in the MIMO uplink channel that varying the detection order of the users using a sorted QR decomposition (SQRD) results in improved detection efficiency by maximizing the diagonal coefficients of $\mathbf{R}$ \cite{husmann_flexcore_2017,fcsd'08}. However, employing such a scheme in the downlink would first necessitate the computation of the channel pseudoinverse, which has a complexity of $\mathcal{O}(N_R^2 N_T)$. Since the SQRD sequentially follows pseudo-inverse operation, this drastically increases the processing latency of the precoder. 

Following on from \cite{husmann'18} which directly decomposes the channel matrix $\mathbf{H}$ as $\mathbf{H}=\mathbf{RQ}$ without explicit computation of $\mathbf{H}^\dagger$, the precoding matrix can be determined by
\begin{align}
	\mathbf{H}^\dagger = \mathbf{H}^H(\mathbf{HH^H})^{-1} &= \mathbf{Q}^H\mathbf{R}^H(\mathbf{R}\mathbf{Q}\mathbf{Q}^H\mathbf{R}^H)^{-1}\\ \notag
	&=\mathbf{Q}^H\mathbf{R}^H (\mathbf{R}\mathbf{R}^H)^{-1} \label{eq:viper_rq1} \\ \notag &= \mathbf{Q}^H \mathbf{R}^{-1}
\end{align}
However, when the number of users exceeds the AP antennas, i.e., $N_T$ $<$ $N_R$, the matrix $\mathbf{HH}^H$ becomes rank-deficient, which implies it is not invertible and the precoding matrix cannot be computed in such cases. 
ViPer NL-COMM solves this problem by employing the sorted RQ decomposition of the Tikhonov regularized channel matrix $\bar{\mathbf{H}}$ \cite{chathu'19}, as follows.
\begin{align}
    \bar{\mathbf{H}} = \begin{bmatrix}
        \mathbf{H} & \lambda \mathbf{I}_{N_R}
    \end{bmatrix} &= \bar{\mathbf{R}}\bar{\mathbf{Q}}
\end{align}
where $\lambda=\tfrac{\sigma}{\textnormal{E}|s_l|}$, $\bar{\mathbf{Q}}$ is an $N_R\times(N_R+N_T)$ unitary matrix and $\bar{\mathbf{R}}$ is an $N_R\times N_R$ upper triangular matrix, and $E[s_l]$ represents the average energy of a symbol in the QAM constellation. This yields a full rank $\bar{\mathbf{H}}\bar{\mathbf{H}}^H$ matrix for which the inverse exists, facilitating the computation of the precoding matrix even in overloaded scenarios, as
\begin{align}
	\bar{\mathbf{H}}^\dagger = \bar{\mathbf{H}}^H(\bar{\mathbf{H}}\bar{\mathbf{H}}^H)^{-1} &= \bar{\mathbf{Q}}^H\bar{\mathbf{R}}^H(\bar{\mathbf{R}}\bar{\mathbf{Q}}\bar{\mathbf{Q}}^H\bar{\mathbf{R}}^H)^{-1}\\
	&=\bar{\mathbf{Q}}^H\bar{\mathbf{R}}^H (\bar{\mathbf{R}}\bar{\mathbf{R}}^H)^{-1} \label{eq:viper_rq2} \\ \notag &= \bar{\mathbf{Q}}^H \bar{\mathbf{R}}^{-1}
\end{align}
\noindent This Tikhonov regularization is equivalent to RZF precoding: with $\lambda>0$ the resulting pseudoinverse corresponds to $\mathbf{H}^H(\mathbf{H}\mathbf{H}^H+\lambda^2\mathbf{I})^{-1}$, whereas $\lambda\!\to\!0$ corresponds to ZF.
This regularization incorporates the MMSE criterion in designing the improved sorted RQ decomposition, whose pseudocode is described in Algorithm 1. It can be observed that in contrast to the SQRD, the proposed method permutes the rows of $\bar{\mathbf{H}}$. The $\bar{\mathbf{R}}$ and $\bar{\mathbf{Q}}$ are updated iteratively, and the channel ordering is captured in the permutation matrix $\bar{\mathbf{P}}$. It must be noted that $\bar{\mathbf{R}}$ is not directly related to the precoding matrix, hence it must be inverted to compute $\bar{\mathbf{H}}^\dagger$. Although triangular matrix inversion has a lower complexity (i.e., $\mathcal{O}(N_R^2)$) than $\mathcal{O}(N_R^3)$ in general $N_R\times N_R$ matrix inversion, it still adds to the resource burden and processing latency of the system.   

\begin{algorithm}[t]
	\caption{Improved sorted RQ decomposition}
	\label{alg:AlgI}
	\begin{algorithmic}[1]
		\Procedure{ISRQD} { ${\mathbf{H}}, \lambda$}
		\State \textbf{Initialization}  ${\bar{\mathbf{Q}}}\leftarrow\begin{bmatrix}
		    {\mathbf{H}} & \lambda\mathbf{I}_{N_R}
		\end{bmatrix}$, $\bar{\mathbf{R}}=\mathbf{0}$, $\bar{\mathbf{P}}=\mathbf{I}$
		\For{$i=1,\cdots,N_R$}
		\State $\textbf{norm}_i=\|\bar{\mathbf{q}_j}\|^2$, where $\bar{\mathbf{q}_j}$ is the $i^\textnormal{th}$ row of $\bar{\mathbf{Q}}$
        \EndFor
		\For{$i=1,\cdots,N_R$}
		\State $p_i=\arg\min_{j=1,\cdots,N_R} \textbf{norm}_j$
        
		\State Swap rows $i$ and $p_i$ in $\bar{\mathbf{Q}}, \bar{\mathbf{R}}, \bar{\mathbf{P}}$ and $\textbf{norm}$

        \State $\bar{{R}}_{i,i}=\sqrt{\textbf{norm}_i}$

        \For{$j=i+1,\cdots,N_R$}
        \State $\bar{{R}}_{j,i}=\bar{\mathbf{q}}\cdot \bar{\mathbf{q}}_i^H$
        \State $\bar{\mathbf{q}}_j=\bar{\mathbf{q}}_j-\bar{{R}}_{j,i}\cdot \bar{\mathbf{q}}_i$
        \State $\textbf{norm}_j=\textbf{norm}_j-|\bar{{R}}_{j,i}|^2$
        \EndFor
		\EndFor
		\State \textbf{return} $\bar{\mathbf{Q}},\bar{\mathbf{R}}, \bar{\mathbf{P}}$
		\EndProcedure
	\end{algorithmic}
\end{algorithm}

This is where the regularization subtly leverages the improved SRQD algorithm to directly obtain 
$\bar{\mathbf{R}}^{-1}$, as shown below.

\begin{align}
\label{eq:invr}
	\bar{\mathbf{H}} = \begin{bmatrix}
        \mathbf{H} & \lambda \mathbf{I}_{N_R}
    \end{bmatrix} &= \bar{\mathbf{R}}\bar{\mathbf{Q}}   \\ \notag
&=\bar{\mathbf{R}}\begin{bmatrix}
     \bar{\mathbf{Q}}_1 & \bar{\mathbf{Q}}_2  
    \end{bmatrix}\\ \notag
    &=\begin{bmatrix}
      \bar{\mathbf{R}}  \bar{\mathbf{Q}}_1 & \bar{\mathbf{R}}  \bar{\mathbf{Q}}_2
    \end{bmatrix} 
\end{align}
where $\bar{\mathbf{Q}}_1$ and $\bar{\mathbf{Q}}_2$ are obtained by partitioning $\bar{\mathbf{Q}}$.

From equation \eqref{eq:invr}, since $\lambda \mathbf{I}_{N_R}=\bar{\mathbf{R}}\bar{\mathbf{Q}}_2$, $\bar{\mathbf{R}}^{-1}$ can be expressed as 
\begin{equation}
    \bar{{\mathbf{R}}}^{-1}=\dfrac{1}{\lambda} \bar{\mathbf{Q}}_2
\end{equation}
Thus, the proposed sorted RQ decomposition in ViPer NL-COMM delivers substantial hardware and latency savings by eliminating explicit matrix inversion. In practice, this removes high-cost divider networks and large control trees, enabling a fully pipelined datapath with lower critical path delay. Importantly, it allows for the first time VP based precoding principles \cite{husmann'18} to be applicable in overloaded scenarios where the number of transmitted streams exceed the number of base-station antennas.

\subsubsection{Identification of Most Promising Perturbation Vectors}
ViPer NL-COMM introduces the ability to identify the most promising perturbation vectors (i.e., potential solutions to Eq. \eqref{eq_qrse}) based only on the regularized channel state information (i.e., $\bar{\mathbf{H}}$), and before any data is available. Considering $\tilde{\mathbf{R}} = \bar{\mathbf{R}}^{-1}$, these most promising perturbation vectors are identified as SE tree paths based on a novel metric that depends on the entries of $\tilde{\mathbf{R}}$. In this direction, a position vector denoted as $\mathbf{p}$ can be introduced to fully describe a specific tree path in terms of the respective node indices in each layer that are ordered according to the increasing PEDs. The absolute real and imaginary magnitudes of the entries in the perturbation vector $\mathbf{t}$ are assumed to be smaller or equal to $\mathcal{B}$, with $\mathcal{B}$ being a positive integer. This results in the branching factor of the SE tree being $|B|=(2\cdot \mathcal{B}+1)^2$, thus limiting the entries of $\mathbf{p}$ ($p_l \in \{1,\ldots,|B|\}$). Unlike the uplink \cite{nik'18}, deriving a metric based only on the channel that quantifies the likelihood of each $N_R$-dimensional position vector $\mathbf{p}$ to be $\mathbf{t}$ is extremely challenging. In this work we use $B=1$, which yields $9$ candidate perturbations per symbol (the zero perturbation plus the eight nearest lattice offsets in the real--imaginary plane). Larger bounds provide diminishing BER gains for substantially higher search complexity.

Instead, ViPer NL-COMM approximates a lower bound of $d(\mathbf{t}_1)$ corresponding to each $\mathbf{p}$ for all possible information vectors \cite{husmann'18}. The derivation of the lower bound is undertaken by first deriving a lower bound for the contribution of each node to $d(\mathbf{t}_1)$. In this direction, consider that $\delta_{\mathbf{p}}(l)$ represents the contribution of the node on the $l^\textnormal{th}$ level of the tree as

\begin{equation}
    \delta_{\mathbf{p}}(l)=|\tilde{R}_{l,l}|^2 \tau^2\, |\hat{s}_l - t_l|^2
\end{equation}

where
\begin{equation}
 \hat{s}_l=\frac{ \tilde{s}_l-\sum _{k=l+1}^{N_R} \tilde{{R}}_{l,k}\tau t_k}{\tilde{{R}}_{l,l}\tau},
\end{equation}
and $t_l$ is the perturbation symbol corresponding to the $l^\textnormal{th}$ level of the tree. Ideally, $t_l$ can be the same as $\hat{s}_l$, assuming that $t_l$ is the closest lattice point to $\hat{s}_l$, which would indicate that the node's contribution to the PED $\delta_\mathbf{p}(l)=0$.  However, $t_l$ may not be the closest lattice point to $\hat{s}_l$,  in order for  $\mathbf{t}$ to solve the joint minimization problem in Eq. 9, and as a result  $\delta_\mathbf{p}(l)>0$. Then,
assuming that $t_{1^\textnormal{st}}$ represents the perturbation symbol nearest to $\hat{s}_l$, a lower bound can be expressed as 
\begin{equation}
\delta_\mathbf{p}(l) \ge |\tilde{R}_{l,l}|^2 \tau^2 \dfrac{| t_{1^\textnormal{st}}-t_l|^2}{2}    
\end{equation}
It was shown in \cite{husmann_flexcore_2017} that the distance between a lattice point and its $p_l^\textnormal{th}$ nearest neighbour scales with $\sqrt{p_l}$. This enables a further approximation in the individual lower bound for $\delta_\mathbf{p}(l)$ as follows
\begin{equation}
    \delta_\mathbf{p}(l) \ge |\tilde{R}_{l,l}|^2 \tau^2 \dfrac{(\sqrt{p_l-1})^2}{2},       
\end{equation}

\noindent which provides a lower bound that does not rely on the data but only the channel state information. By omitting the constant terms, the overall lower bound can be approximated as
\begin{equation}
    d(\mathbf{t}_1)=\sum_{l=1}^{N_R}\delta_\mathbf{p}(l) \ge \sum_{l=1}^{N_R} |\tilde{R}_{l,l}|^2(p_l-1),
\end{equation}
and yields the final metric, which is denoted as the Metric of Promise ($\mathcal{M}$) and can be defined as follows
\begin{equation}
    \mathcal{M}(\mathbf{p})=\sum_{l=1}^{N_R} |\tilde{R}_{l,l}|^2(p_l-1).
    \label{eq_mop}
\end{equation}

With $\mathcal{M}$ in place, identifying the $K$ most promising perturbation vectors becomes a tree-search problem, where the objective is to find the $K$ vectors attaining the smallest MoP values (i.e., MoP minimization).
We adopt a breadth-first \emph{K-Best} search that uses the partial MoPs, as a minimization pruning metric. Starting at $l=1$, each surviving partial path is expanded to all $|B|$ children, we then keep the $K$ children with the smallest $\mathrm{MoP}_{1:l} (\mathbf{p}_{1:l}) \triangleq \sum_{i=1}^{l} |\tilde{R}_{i,i}|^2\,(p_i-1)$ and prune the rest. This repeats for $l=1,\dots,N_R$, yielding $K$ full-depth candidates $\{\mathbf{p}^{(k)}\}_{k=1}^{K}$.
Operationally, at each level, we (i) compute the $\;K|B|\;$ partial metrics for all children of the current $K$ survivors, (ii) rank them by
$\mathrm{MoP}_{1:l}(\mathbf{p}_{1:l})$,  and (iii) retain only the $K$ smallest to expand further at the following tree level. 
The ultimate decision is made by evaluating the true PED for each surviving $\mathbf{p}^{(k)}$ In particular, the most promising perturbation vectors are demapped to actual perturbations based on $\hat s_l$ (Eq. 17) by using a predefined ordering \cite{chathu'19}. Finally, the perturbation with the minimum PED is selected as $\mathbf{t}$
to produce the precoded signal to be transmitted as
\begin{equation}
   \mathbf{v}_\textnormal{precoded}=\sqrt{\dfrac{P_T}{\gamma_\textnormal{VP}}} \mathbf{H}^\dagger (\mathbf{v}-\tau\mathbf{t})
\end{equation}

Algorithm \ref{alg:viper_end2end_simple}, presents in high level the end-to-end ViPer-NLCOMM process. Its worth mentioning here that
in the \emph{DEMAP} process, each candidate position index $p_l$ at level $l$ is converted into an actual perturbation symbol $t_l$ by selecting the $p_l$-th element from a predefined neighbour ordering around the effective point $\hat{s}_l$. Rather than exhaustively evaluating and sorting Euclidean distances to all possible perturbation symbols, we exploit the regular grid structure and symmetry of the underlying complex lattice. Specifically, we first translate $\hat{s}_l$ to the coordinate system of its nearest lattice center and determine its relative position within the fundamental square. This region index is then used to address a small lookup table (LUT) that stores a frequently occurring ordered list of neighbour offsets; the $p_l$-th entry of this list provides the perturbation symbol for that tree node. This approach follows the LUT-based symbol enumeration strategy of FlexCore \cite{husmann_flexcore_2017} (which avoids unnecessary distance computations) and is adapted here to the perturbation alphabet used by ViPer NL-COMM. Full details of LUT construction and symmetry-based reuse are provided in detail in our previous publication~\cite{husmann_flexcore_2017}.


\begin{algorithm}[t]
\caption{End-to-end ViPer NL-COMM}
\label{alg:viper_end2end_simple}
\begin{algorithmic}[1]
\Require $\mathbf{H}\!\in\!\mathbb{C}^{N_R\times N_T}$, noise std.\ $\sigma$, data vector $\mathbf{v}\!\in\!\mathbb{C}^{N_R}$, constellation scaling $\tau$, transmit power $P_T$, num.\ candidates $K$

\Statex \textbf{Regularize \& factorize:}
\State $\lambda = \sigma / \mathbb{E}[\,|s_l|\,]$
\State $\bar{\mathbf{H}} = \big[\mathbf{H}\;\;\lambda\mathbf{I}_{N_R}\big]$
\State $(\bar{\mathbf{R}},\bar{\mathbf{Q}},\bar{\mathbf{P}}) \gets \textsc{ISRQD}(\bar{\mathbf{H}})$
\State $\tilde{\mathbf{R}} \gets \bar{\mathbf{R}}^{-1}$ \Comment{obtained implicitly (e.g., via $\tilde{\mathbf{R}}=\bar{\mathbf{Q}}_2/\lambda$)}

\Statex \textbf{Most Promising Perturbation Vector Identification (channel-only):}
\State $\{\mathbf{p}^{(k)}\}_{k=1}^{K} \gets \textsc{MoP-Search}(\tilde{\mathbf{R}},K)$ \Comment{e.g., K-Best}

\Statex \textbf{Parallel post-processing over $K$ candidates (data-dependent):}
\ForAll{$k\in\{1,\dots,K\}$ \textbf{in parallel}}
    \State $\mathbf{t}^{(k)} \gets \textsc{Demap}\!\big(\mathbf{p}^{(k)},\tilde{\mathbf{R}},\mathbf{v},\tau\big)$
    \Comment{\small LUT-based}
    \State $\mathbf{w}^{(k)} \gets \bar{\mathbf{Q}}^{H}\tilde{\mathbf{R}}\big(\mathbf{v}-\tau\mathbf{t}^{(k)}\big)$
    \State $d^{(k)} \gets \|\mathbf{w}^{(k)}\|^2$ \Comment{\small PED for candidate $k$}
\EndFor
\State $k^\star = \arg\min_k d^{(k)}$
\State $\mathbf{w} \gets \mathbf{w}^{(k^\star)}$;\;\; $\gamma_{\mathrm{VP}} = d^{(k^\star)}$
\State $\mathbf{v}_{\textnormal{precoded}} = \sqrt{\dfrac{P_T}{\gamma_{\mathrm{VP}}}}\;\mathbf{w}$
\State \Return $\mathbf{v}_{\textnormal{precoded}}$
\end{algorithmic}
\end{algorithm}

\subsection{ViPer NL-COMM Evaluation}

\begin{table}[ht]
\centering
\caption{Simulation Parameters and System Configuration.}
\label{tab:sim_config}
\renewcommand{\arraystretch}{1.2}
\resizebox{\linewidth}{!}{
\begin{tabular}{l l}
\hline
\textbf{Parameter} & \textbf{Value} \\
\hline
\multicolumn{2}{c}{\textit{OFDM Numerology \& Frame Structure}} \\
\textbf{Waveform / SCS} & CP-OFDM / 30 kHz \\
\textbf{Slot Duration} & 0.5 ms (14 OFDM symbols) \\
\textbf{Simulated Bandwidth} & 8 Resource Blocks / 96 subcarriers$^{\mathrm{a}}$ \\
\textbf{Channel Update} & Once per slot \\
\hline
\multicolumn{2}{c}{\textit{Transmission \& Channel}} \\
\textbf{Tx/Rx Antennas} & $N_T\in[4,16]$, $N_R\in[4,16]$ \\
\textbf{Channel Model} & i.i.d.\ Rayleigh / 3GPP CDL-B (5 km/h) \\
\textbf{Modulation} & QPSK, 16-QAM, 64-QAM, 256-QAM \\
\textbf{Coding Scheme} & LDPC ($N=1944$, $R \in \{1/2, 2/3, 5/6\}$) \\
\textbf{SNR Definition} & $P_T/\sigma^2$ (at Receiver Input) \\
\hline
\multicolumn{2}{l}{$^{\mathrm{a}}$\footnotesize{Default configuration unless explicitly stated otherwise.}}
\end{tabular}}
\end{table}

In this subsection, we analyze the performance of the ViPer NL-COMM algorithm in various MIMO configurations. All link-level simulations were performed in MATLAB R2024b. We utilized i.i.d.\ Rayleigh channels. We evaluate different MIMO sizes, quadrature amplitude modulation (QAM) orders, low-density parity check (LDPC) code rates, and signal-to-noise ratios (SNRs). Coded simulations employ LDPC with a block length of 1944 bits.

\begin{figure}[h]
    \centering
    
    \includegraphics[width=0.8\linewidth]{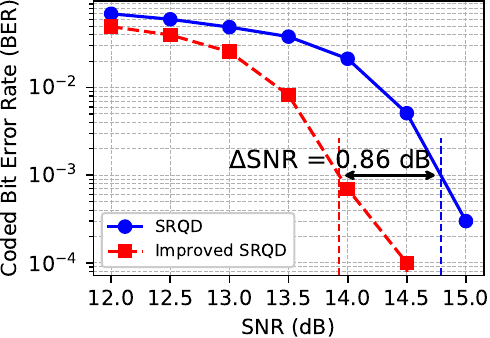}
    \caption{Coded bit error rate performance comparison of ViPer \cite{husmann'18} and ViPer-NLCOMM, assuming 8 $\times$ 8 Rayleigh fading channels with 64-QAM modulation and LDPC code rate 5/6.}
    \label{fig:regViPer}
\end{figure}

First, we examine the impact of the improved SRQD on the precoding performance of ViPer NL-COMM. In addition to enabling precoding in overloaded scenarios and simplifying the hardware structure, Fig. \ref{fig:regViPer} shows how the improved SRQD improves the coded bit error rate (BER) of ViPer by roughly 0.9 dB in an 8$\times$8 MIMO system. This is attributed to the improved channel conditioning enabled by this approach, which results in stronger channel components being processed first. Consequently, this improves the performance of the subsequent tree layers in the identifying the near-optimal perturbation vectors by facilitating early pruning of the false branches.

\begin{figure}[ht]
    \centering
    \begin{subfigure}{.5\textwidth}
    \centering
        \includegraphics[scale=.75]{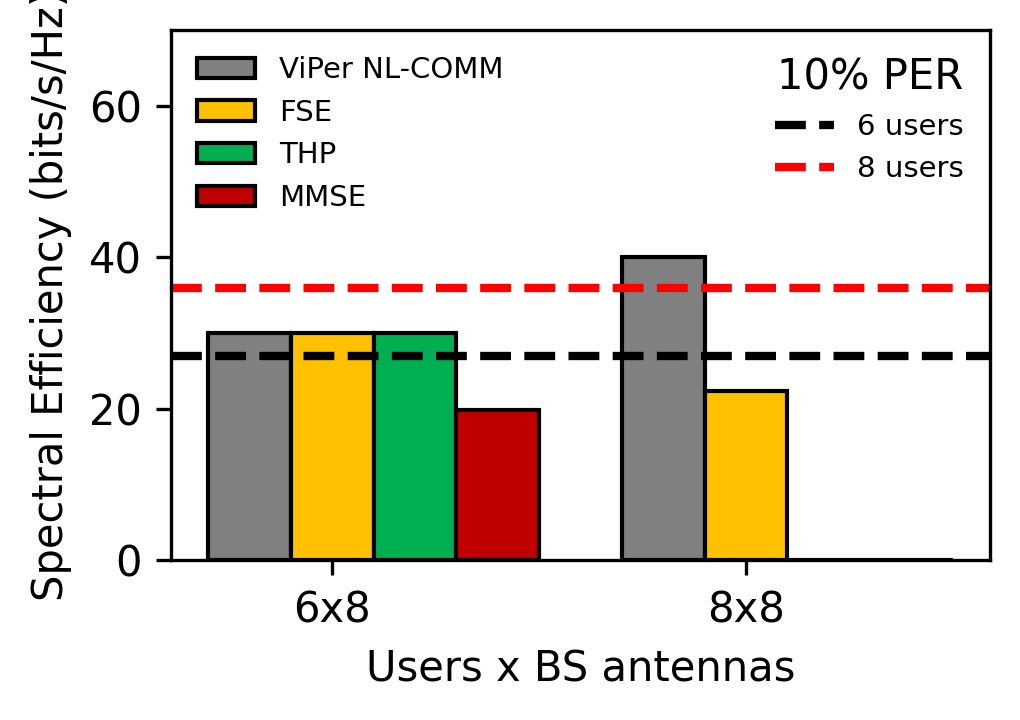}
        \caption{}
        \label{fig:itput}
    \end{subfigure}%

    \begin{subfigure}{.5\textwidth}
    \centering
        \includegraphics[scale=.75]{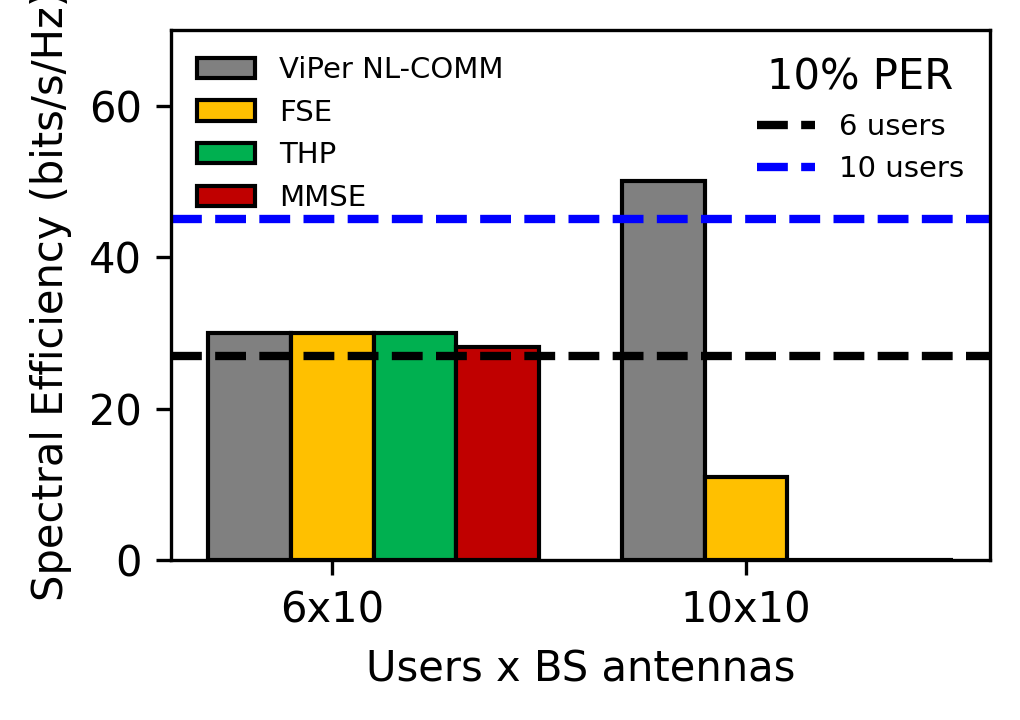}
        \caption{}
        \label{fig:itput2}
    \end{subfigure}
    \caption{Spectral efficiency comparison of MMSE, THP, FSE, and ViPer for (a) 8 AP antennas and (b) 10 AP antennas, assuming Rayleigh fading channels with 64-QAM modulation and LDPC code rate 5/6 at an intermediate SNR of 17 dB. Dashed lines represent the 10\% PER boundaries for different number of concurrently transmitted streams.}
    \label{fig:tput_overall}
\end{figure}

In Fig. \ref{fig:tput_overall}, we specifically show the performance improvement achieved by ViPer NL-COMM over existing linear (i.e., MMSE) and non-linear (i.e., THP and FSE) precoding algorithms. 
This simulation employs 64-QAM and an LDPC code rate of 5/6, assuming an intermediate SNR of 17 dB. 
%
Fig. \ref{fig:itput} shows that ViPer NL-COMM, FSE and THP, all meet the targeted 10\% PER  when 6 concurrent users are supported by an 8-antenna AP, while linear MMSE does not. However, while ViPer NL-COMM can support 8 concurrent users using the same 8-antenna AP, FSE falls below the threshold for even 6 users, while  THP and MMSE are unable to support even a single stream. Thus, for the same 8-antenna AP, ViPer NL-COMM achieves 30\% spectral efficiency improvement over THP and FSE with the same complexity, and 100\% over MMSE. Fig. \ref{fig:itput2} shows that when 6 concurrent users are supported by a 10-antenna AP, all algorithms including MMSE perform above the 10\% PER threshold. However, only ViPer NL-COMM supports 10 users using the same number of AP antennas, whereas FSE performs 4$\times$ worse, and the others fail completely. It is worth mentioning the ViPer NL-COMM for the given evaluation processes 8 most promising path solutions.

\begin{figure*}
	\centering
	\includegraphics[scale=.5]{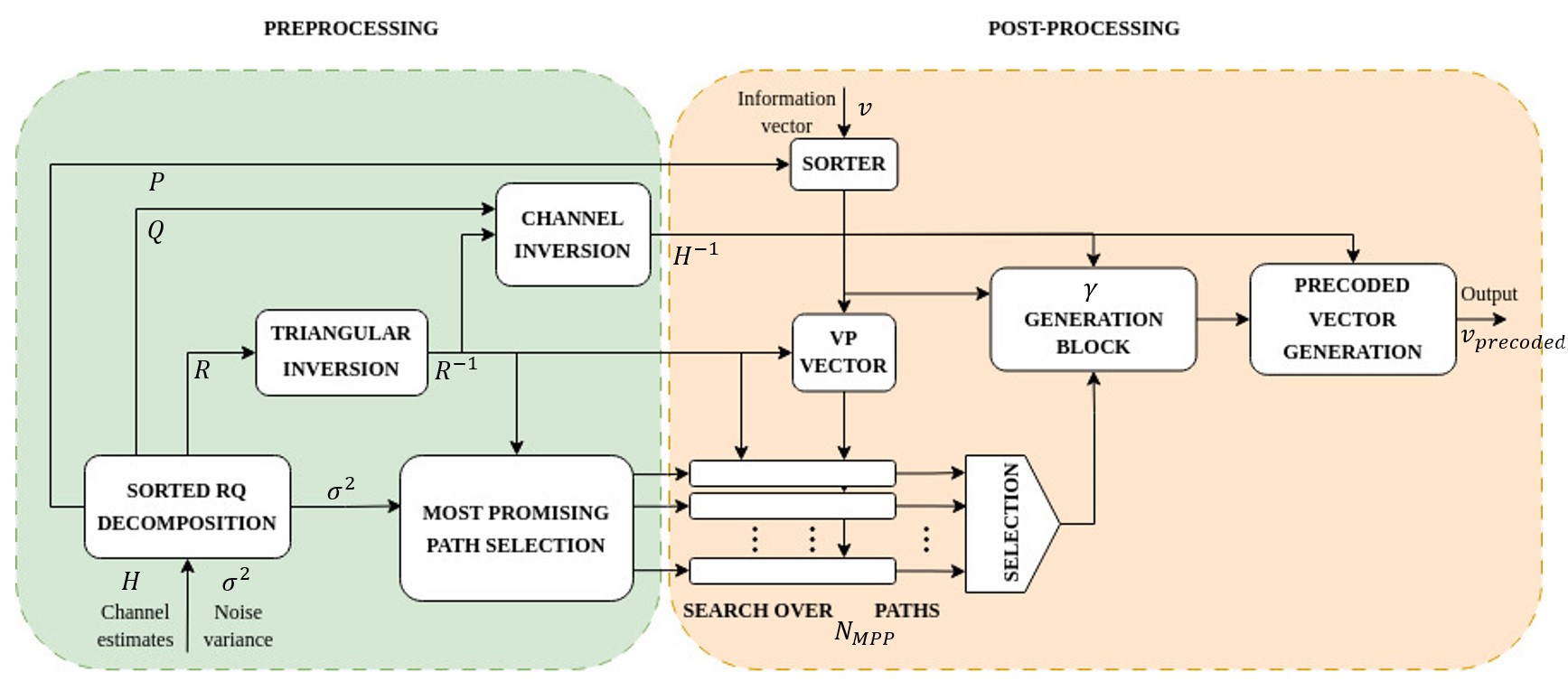}
	\caption{Top-level architecture of the ViPer NL-COMM based non-linear precoder.}
	\label{fig:arch_block}
\end{figure*}
\begin{figure}[h]
    \centering
    \includegraphics[width=.87\linewidth]{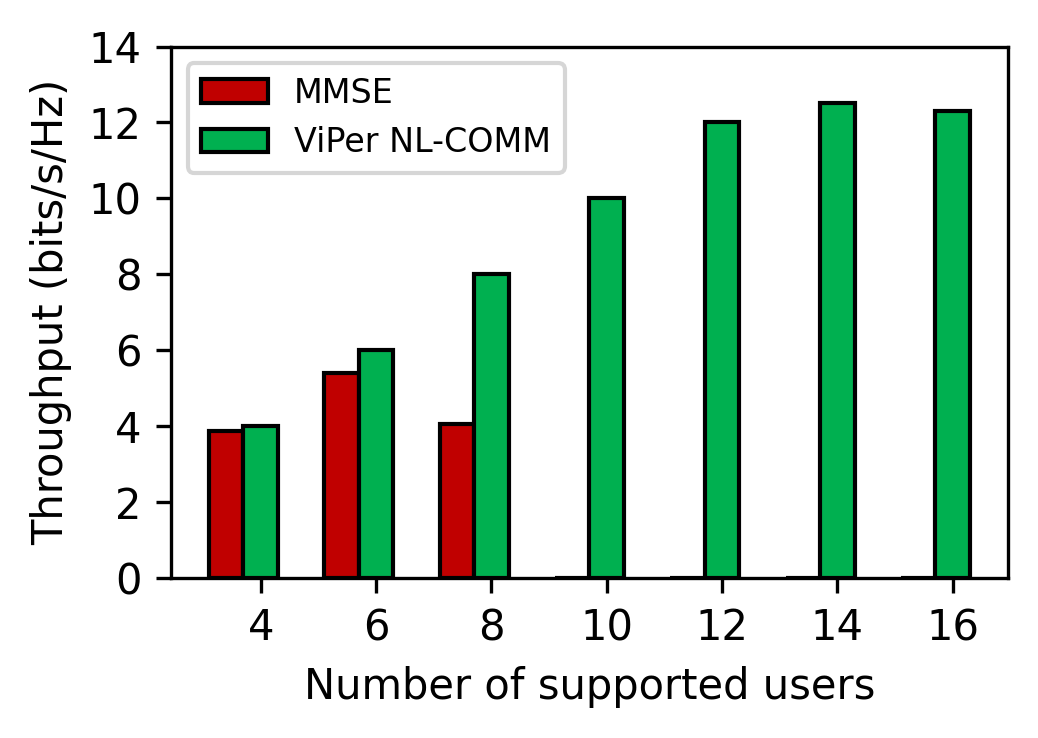}
    \caption{Spectral efficiency comparison of linear (MMSE) and non-linear (ViPer) precoding for an 8-antenna AP serving single-antenna low-rate users employing 4-QAM and 1/2 rate LDPC codes with 1944 block length at an SNR of 17 dB.}
    \label{fig:viper_overloaded}
\end{figure}

Moreover, as discussed earlier, ViPer NL-COMM is capable of fully exploiting the available channel capacity, supporting even more users than the number of AP antennas. This ability to deliver overloading gains is particularly relevant for massive connectivity use-cases where the transmitted streams are of low rate, commonly in Internet of Things (IoT) scenarios. More specifically, Fig. \ref{fig:viper_overloaded} shows that ViPer can consistently support up to twice the number of users as AP antennas in such overloaded scenarios, while the MMSE approach. 

\begin{figure}
    \centering
    \includegraphics[]{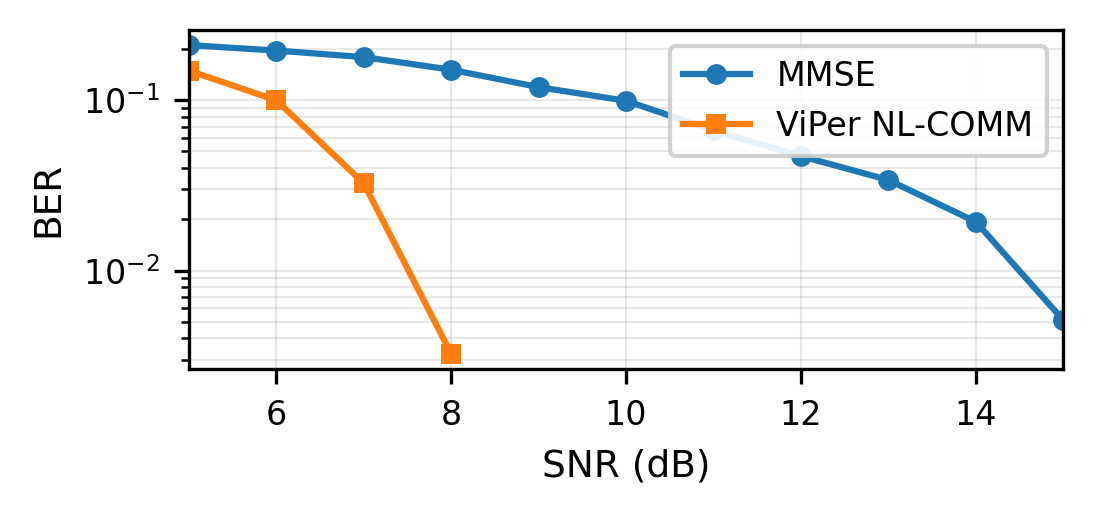}
    \caption{Coded BER performance for an 8x8 MU-MIMO MMSE vs ViPer NL-COMM under 3GPP CDL-B (5~km/h) using QPSK and 0.5 code rate.}    \label{fig:cdl_mmse_viper}
\end{figure}

While the above results assume i.i.d. Rayleigh fading, we also evaluated ViPer NL-COMM under a standardized multipath channel models. Indicatively, Fig.~\ref{fig:cdl_mmse_viper} compares MMSE and ViPer NL-COMM coded error-rate performance using the 3GPP CDL-B profile with user mobility of 5~km/h (QPSK). The observed trend is consistent with the Rayleigh-fading results.

Finnaly, in terms of computational complexity the matrix inversion required for MMSE and the Sorted RQ Decomposition (SRQD) required for ViPer NL-COMM is asymptotically the same, with both operations scaling on the order of $2N_rN_t^2$ complex multiplications. These operations are performed on the channel matrix whenever channel estimates are updated. The practical implementation costs differ significantly since MMSE requires explicit matrix inversion to compute the precoding matrix. In hardware, this typically necessitateess computationally expensive divider networks. ViPer NL-COMM replaces explicit inversion with a SRQD. As detailed in Section III-A, our design avoids high-cost division operations in the critical path by obtaining the inverse factor implicitly from the unitary matrix decomposition. This results in a more hardware-efficient architecture.
For processing per transmitted symbol vector, the costs diverge: MMSE requires $N_rN_t$ complex multiplications per transmitted symbol vector, whereas ViPer NL-COMM requires $N_rN_t + N_t^2 + KN_t(1 + (N_t+1)/2)$.

\section{ViPer NL-COMM Architecture Design}
This section discusses the architectural design of ViPer NL-COMM, from a high-level overview to the detailed descriptions of the fundamental processing blocks and the design considerations. 

\begin{figure}[h]
    \centering
    \includegraphics[]{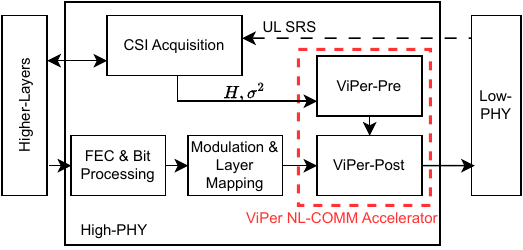}
    \caption{System-level placement of the ViPer NL-COMM accelerator within the PHY stack. The dashed box marks the HW-accelerated functionalities.}
    \label{fig:highlevel}
\end{figure}

\subsection{Placement within Protocol Stack}
Figure~\ref{fig:highlevel} summarizes where ViPer NL-COMM integrates with the 5G NR PHY.  The offloaded blocks reside within High-PHY and are partitioned into a preprocessing stage (ViPer-Pre), which consumes new channel/noise estimates and produces the sorted-\(RQ\) artifacts required for the precoding, and a postprocessing stage (ViPer-Post), which generates the final precoding vectors per subcarrier/layer for Low-PHY to consume. High-PHY retains the forward-error-correction (FEC) chain, as well as functionalities such as modulation and layer mapping. The detailed architecture of the accelerated blocks follows in the subsequent sections.

Its worth mentioning here that in a standards-compliant system, a UE would advertise support for VP during RRC connection establishment as part of the UE Capability Information procedure. This enables the gNB scheduler to pair and allocate VP-compatible UEs for nonlinear-precoded transmissions. During data transmission, the DCI would simply require a minor extension, such as introducing an additional flag bit in DCI Format 1.1 (38.212, Table 7.3.1-102), to indicate whether the scheduled PDSCH uses a perturbed (VP) precoder.

\subsection{Architecture Overview}\label{ss_arch}
The high-level architecture of the key steps involved in generating the ViPer precoding vectors is shown in Fig. \ref{fig:arch_block}, which can be split into preprocessing and postprocessing. The preprocessing functional blocks encompass the essential steps to be performed whenever new channel state information (CSI) and/or noise estimates are available. The frequency of preprocessing operations depends on the coherent interval of the specific MIMO environment. For instance, it can range from one preprocessing execution every couple of slot durations for slowly changing environments to once per 5G NR symbol in case of highly mobile users \cite{vehicular_2024}. For the sake of simplicity, the evaluations in this work assume that preprocessing needs to be carried out on each slot.

The sorted RQ block is the most computationally challenging task owing to its complexity of $\mathcal{O}(N_T N_R^2)$. Though the sorted RQ algorithm vastly differs from the traditional SQRD operation, this work considers the need for efficiently reusing the corresponding SQRD structure from the preprocessing phase of the NL-COMM detector \cite{nik'18} in the uplink slot, allowing symmetric uplink and downlink preprocessing implementation, and introduces the following optimization. 

Assuming $\mathbf{H}^T$ is passed to the SQRD structure of the detector, the corresponding $\tilde{\mathbf{Q}}$ and $\tilde{\mathbf{R}}$ outputs can be rearranged to yield the sorted RQ decomposition outputs $\mathbf{Q}$ and $\mathbf{R}$, as shown below.
\begin{equation}
	\mathbf{Q} = 
	\begin{bmatrix}
		\tilde{q}_{1N_R} & \tilde{q}_{2N_R} & \cdots & \tilde{q}_{N_TN_R} \\
		\vdots & \vdots & \ddots & \vdots \\
		\tilde{q}_{N_T2} & \tilde{q}_{(m-1)2} & \cdots & \tilde{q}_{12} \\
		
		\tilde{q}_{11} & \tilde{q}_{21} & \cdots & \tilde{q}_{N_T1}
	\end{bmatrix}
	\nonumber
\end{equation}
\begin{equation}
	\mathbf{R} = 
	\begin{bmatrix}
		\tilde{r}_{N_RN_R}  & \cdots & \tilde{r}_{2N_R} & \tilde{r}_{1N_R}\\
		0 & \ddots & \vdots & \vdots \\
		0 & 0 & \tilde{r}_{22} & \tilde{r}_{12} \\
		0 & 0 & 0 & \tilde{r}_{11}
	\end{bmatrix} 
	\nonumber
\end{equation}
It can be observed from Fig. \ref{fig:arch_block} that subsequent processing requires $\mathbf{R}$ to be inverted. Instead of actually performing inversion on R, this work exploits the knowledge that the MMSE regularized SQRD preserves a scaled version of $\mathbf{R}^{-1}$ on the bottom $N_T$ rows of the extended $\mathbf{Q}$ matrix. Hence, a simple descaling operation will yield the desired inverse of $\mathbf{R}$, which considerably saves resources and clock cycles. The SQRD architecture proposed in \cite{thomas'24} enables an interleaved pipelining approach to process blocks of multiple subcarriers in subsequent clock cycles.

The most promising path selection block is tasked with finding the MPP most suitable paths to search for the perturbation vector. That is the MPP paths with the smallest MoP (Eq.~\ref{eq_mop}). Because the latency of path selection block is comparable to the SQRD latency in \cite{thomas'24}, it is spatially multiplied to simultaneously determine the best possible paths for each of the corresponding subcarriers from the sorted RQ block.

The postprocessing block is designed as a fully pipelined architecture that accepts  $\mathbf{R}^{-1}$, ${\mathbf{v}}$ and the MPP inputs from the previous blocks to calculate the corresponding partial Euclidean distances (PEDs) and the symbol vectors in a successive manner. 
First is the VP vector generation block is responsible for producing the $\tilde{\mathbf{y}}$ vectors that transform the information vector $\mathbf{v}$ to perform the integer lattice search, according to Eq.~\ref{eq:se}. The permutation matrix is used to apply the effects of sorted RQ while generating $\tilde{\mathbf{y}}$ as
\begin{equation}
	\tilde{\mathbf{y}}=\mathbf{R}^{-1}\mathbf{P}\mathbf{v}
\end{equation}

The processing of the vectors initiates from the bottom layer (i.e., $N_R$) with each layer's outputs being used to feed the upper layer inputs, and continues till the topmost layer is processed. It should be noted that this block avoids explicit division by the corresponding diagonal elements of $\mathbf{R}^{-1}$  at each layer with the available diagonal elements of $\mathbf{R}$ from the sorted RQ block.

The channel inverse computation block avoids the direct inversion of $\mathbf{H}$ and exploits the computed $\mathbf{Q}$ and $\mathbf{R}^{-1}$ matrices. However, explicit construction of $\mathbf{H}^{\dagger}=\mathbf{Q}^H\mathbf{R}^{-1}$ is also bypassed as its application in the $\gamma$ generation block can be decomposed into successive transformations of the precoding vector in a highly pipelined manner, as follows. 
\begin{align}
\mathbf{w}&=\mathbf{Q}^H\{\mathbf{R}^{-1}(\mathbf{v}-\tau\tilde{\mathbf{t}}_{\min})\} \\ \nonumber
&=\{\mathbf{Q}^H\mathbf{z}\} (\because \mathbf{z}=\mathbf{R}^{-1}(\mathbf{v}-\tau\tilde{\mathbf{t}}_{\min}))
\end{align}
Furthermore, $\mathbf{w}$ computations can be reused in the Precoded Vector (PV) generation block to avoid wasteful duplication.

In order to ensure that the maximum transmit power is not exceeded, the $\gamma$ generation block is used to generate appropriate normalization values for each precoded vector as 
\begin{equation}
	\gamma = \|\mathbf{w}\|^2
\end{equation}
It can be observed that the precoding vector has to be multiplied by the reciprocal square root (RSR) of $\gamma$ before transmission. A dedicated RSR unit based on a scaling-less Newton-Raphson architecture is employed to determine this quantity with a high level of precision. Therefore, the operation of the PV generation block is simplified by multiplying the RSR of $\gamma$ with the respective vector $\mathbf{w}$ to obtain the ViPer NL-COMM precoded vector.

\begin{figure}[h]
    \centering
    \includegraphics[width=.9\linewidth]{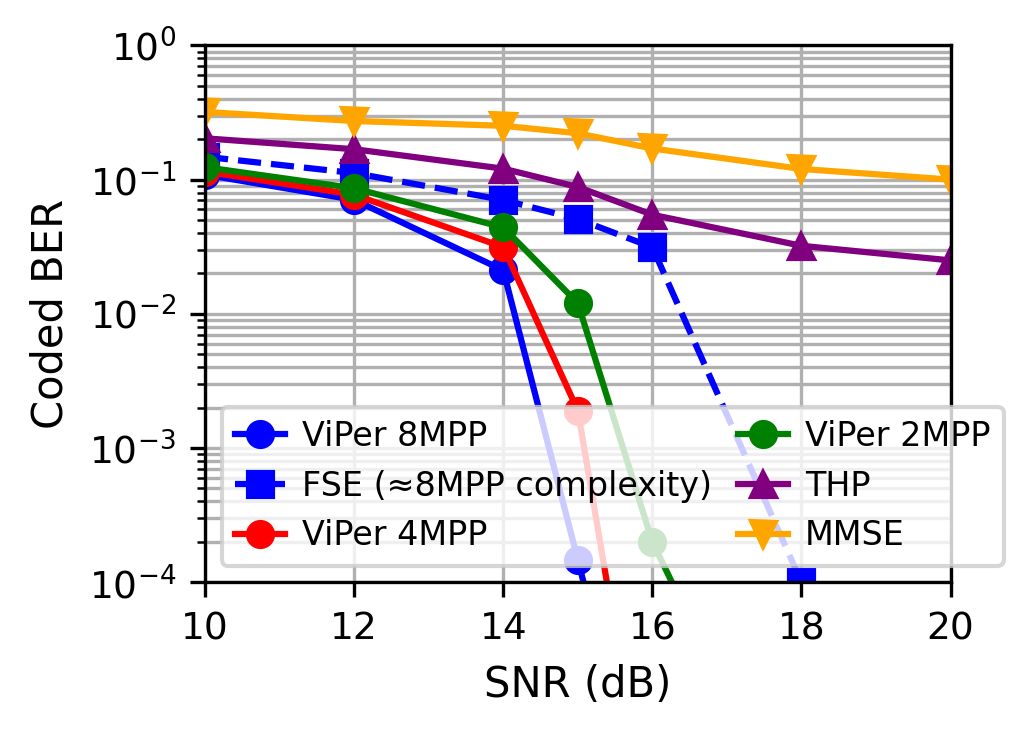}
    \caption{Coded transmission bit error rate (BER) vs SNR for different number of most processing paths (MPPs) in an 8$\times$8 system with 64-QAM and 5/6 rate LDPC codes. Comparing with MMSE, THP, and FSE.}
    \label{fig:npe_graph}
\end{figure}
\subsection{Design Considerations}
The number of most promising paths (MPPs) is a key design parameter in the ViPer NL-COMM implementation, as it affects precoding performance and hardware complexity. Fig. \ref{fig:npe_graph} shows how the bit error rate can be improved as the number of MPPs increases from 2 to 8 for an 8$\times$8 system (64-QAM, 5/6 code rate). For equal hardware complexity, ViPer NL-COMM significantly outperforms FSE and the THP/MMSE baselines. Furthermore, due to the massively parallel nature of ViPer NL-COMM, each path can be processed independently of the other, but this may become impractical due to the resource limitations of the FPGA. 
To address this, we combine spatial and pipeline parallelism across the MPPs, trading off throughput against resource usage. Concretely, we instantiate $N_{\text{PE}}{=}8$ parallel elements (PEs). Each PE processes an arbitrary number of MPPs in a streaming (pipelined) fashion. The resulting latency per precoding vector is $\lceil N_{\text{MPP}} / N_{\text{PU}} \rceil$ clock cycles, while sustaining one MPP result per cycle once the pipeline is filled.

\section{Implementation}
This section discusses the implementation aspects of the proposed architecture, and compares its performance with the existing state-of-the-art solutions. We report the FPGA implementation results of ViPer NL-COMM in terms of computational throughput, latency, resource utilization, power consumption, and discuss aspects related to the PCIe interfacing and the numerical accuracy of the implemented design.

\begin{table*}
	\centering
	\caption{Comparison of the proposed VP precoding accelerator with state-of-art FPGA-based precoders.}
	\renewcommand{\arraystretch}{1.2}
	\resizebox{.9\textwidth}{!}{
		\begin{tabular}{||c|c|c|c|c|c|c|c|c|c||}
			\hline
			\hline
			\multirow{2}{*}{\textbf{Reference}} &\multirow{2}{*}{\textbf{\cite{barrenec'11}}}
			& \multirow{2}{*}{\textbf{\cite{krivo'19}}}
			&\multirow{2}{*}{\textbf{\cite{haq'22}}}
			& \multirow{2}{*}{\textbf{\cite{barren'13}}}
			& \multirow{2}{*}{\textbf{\cite{barren'13}}}
			& \multicolumn{4}{c||}{{\textbf{This Work}}}
			
			\\ 
			
			&
			&
			&
			&
			&
			&  \multicolumn{4}{c||}{{}}

			\\
			
			\hline\hline
			\multirow{2}{*}{\textbf{MIMO system}} &  4$\times$4
			& 6$\times$6
			& 4$\times$4
			& 4$\times$4
			& 4$\times$4
			& 4$\times$4
			& 8$\times$8
            & 12$\times$12
            & 16$\times$16
			\\ 
			
			& 16-QAM
			& 16-APSK
			& 4-QAM
			& 16-QAM
			& 16-QAM
			& 16-QAM
			& 16-QAM
            & 16-QAM
            & 16-QAM
			\\
			\hline
			
			\textbf{FPGA} &
			Virtex-6
			& Kintex-7
			& Kintex-7
			& Virtex-6
			& Virtex-6
			&\multicolumn{2}{c|}{XCVU9P}
            &\multicolumn{2}{c||}{XCU250}
			
			\\ \hline
			
			\textbf{Precoding approach} & VP
			& SLP
			& SLP
			& VP K-Best
			& VP FSE
			&\multicolumn{4}{c||}{ViPer NL-COMM}
			
			\\ \hline

			\textbf{Preprocessing} & Not included
			& Not included
			& Not included
			& Not included
			& Not included
			&\multicolumn{4}{c||}{\textbf{Preprocessing Included}}
			
			\\ \hline
            
			\textbf{LUTs}& 227520
			& 2488
			& 57966
			& 31488
			& 11020
			& 216414
			& 317712
            & 506325
            & 626926
			\\ \hline
			
			\textbf{FFs} &  81907
			& 8076
			& 12562
			& 31488
			& 15744
			& 221001
			& 335702
            & 570320
            & 554963
			\\ \hline
			
			\textbf{DSPs} & 265
			& 72
			& 264
			& 230
			& 230
			& 1921
			& 4192
            & 8352
            & 12102
			\\ \hline
			
			\textbf{Frequency}[MHz] & 97
			& 166
			& 113
			& 345
			& 352
			& 424
			& 410
            & 264
            & 256
			\\  \hline

            \textbf{Throughput}[Gbps] & 2.4
            & 1.33
            & 1.6
            &5.52
            &5.63
            &6.78
            &13.12
            &12.67
            & 16.38
            \\ \hline


			\hline
	\end{tabular}}
	\label{tab:comparison}
\end{table*}

\subsection{Experimental setup and validation}\label{sec:exp_setup}
The designs were synthesized and implemented on an AMD Xilinx VCU118 (Virtex UltraScale+ XCVU9P), using highly parameterized Verilog and VHDL RTL, and hosted in a PCIe Gen3-capable Skylake workstation equipped with an 18-core Intel Core i9-7980XE CPU @ 2.60 GHz.  The implemented design achieves a post-route timing-closed frequency of 428MHz. The ViPer blocks were synthesized with retiming enabled using AMD Vivado.  
The accelerator uses 16-bit fixed-point arithmetic internally (Q8.8)
with functional correctness verified against a floating-point reference (MATLAB) across MIMO sizes and constellations, as discussed in more detail in Section \ref{percision}. Channel models where relevant follow i.i.d.\ Rayleigh fading.
It should be noted that the number of MPPs is selected as 8 for the subsequent evaluations, as it achieves a good trade-off between precoding performance and hardware complexity by using four parallel path processing units, as we discuss later, in an interleaved pipelining fashion.

The reported results assume 30~kHz SCS (500~\textmu s/slot), 14 OFDM symbols/slot with two DMRS (12 data symbols), and 4x4, 8x8, 12x12 and 16x16 MIMO configurations. Modulation orders up to 64-QAM are supported the corresponding LDPC code rates do not affect the ViPer NL-COMM computational latency or throughput. The preprocessing stage that depends on CSI is assumed once per slot. The power estimates and utilization report were obtained with AMD Power Estimator (worst-case static conditions). The processing latency results were obtained from post-synthesis simulations and include all internal pipelining overheads.
It is worth mentioning here that the results in this section focus on computational-time aspects to provide a transferable evaluation across implementation boards, decoupled from board-specific PCIe tuning. A detailed discussion of PCIe interfacing overheads appears in Sec.~\ref{pcie_overhead}.

\subsection{Precision and Numerical Accuracy}\label{percision}
\begin{figure}[h]
    \centering
    \includegraphics[width=.9\linewidth]{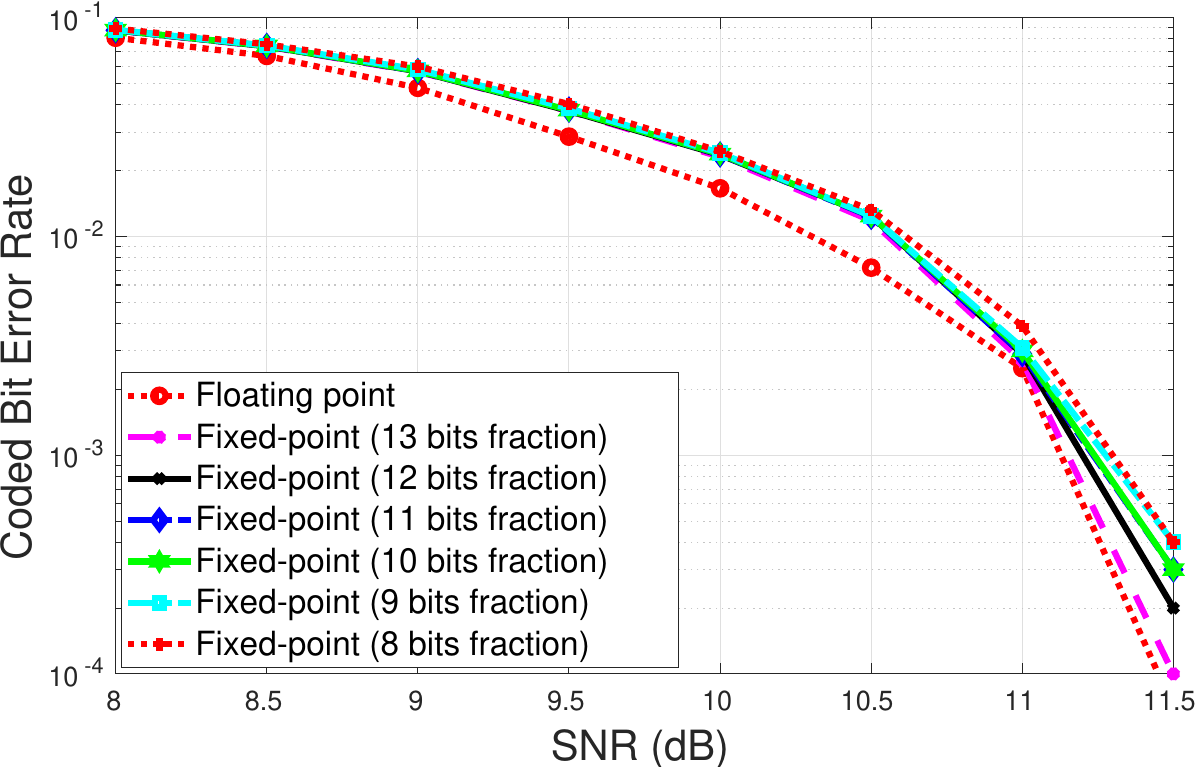}
    \caption{Coded BER vs. SNR for different fractional precision formats compared to the floating-point implementation.}
    \label{fig:fpa_graph}
\end{figure}

The accelerator uses a signed 16-bit fixed-point format to align with common open-source PHY stacks (e.g., OpenAirInterface and srsRAN). We assess the impact of fractional precision by varying the number of fractional bits from 8 to 13 while keeping the total width at 16.
As shown in Fig.~\ref{fig:fpa_graph}, coded BER improves monotonically with fractional precision; at 13 fractional bits, the fixed-point design achieves the best performance among the tested formats relative to the baseline floating-point implementation. Unless otherwise stated, results use 13 fractional bits.


\subsection{FPGA Implementation Results}
Table \ref{tab:comparison} compares the resource utilization of the ViPer NL-COMM implementation with the state-of-the-art FPGA-based designs. It must be noted that the existing designs implement only the post-processing functionality, assuming that the results from the preprocessing (e.g., R, Q matrices) are available in the FPGA. In practice, this substantially limits the achievable real-time capabilities as the transfer times over PCIe become significantly higher. Instead, our proposed design is the first that accelerates the entire precoding chain on the same FPGA device, justifying its relatively higher resource utilization.
%

For a system with $N_R$ users and a constellation of $Q$ symbols, the
throughput for a fully-pipelined architecture can be computed as follows \cite{barren'13}
\begin{equation}
    T=N_R f_{\textnormal{clk}}\log_2 (Q)
    \label{eq:throughput}
\end{equation}
where $f_{\textnormal{clk}}$ is the maximum operating clock frequency of the design. As mentioned earlier, the 4$\times$4 and 8$\times$8 designs with four and eight PEs, respectively, enable a fully-pipelined architecture, achieving the maximum throughput as (\ref{eq:throughput}). If a higher number of paths given by $N_{\textnormal{MPPs}}$ or higher MIMO dimensions need to be accommodated, the processing can be time-multiplexed over multiple PEs by employing a folding factor $N_\textnormal{fold}$, which indicates the throughput achieved in such cases will be
\begin{equation}
    T=N_R f_{\textnormal{clk}}\dfrac{\log_2 (Q)}{N_\textnormal{fold}}
    \label{eq:throughput2}    
\end{equation}
The proposed 4 $\times$ 4 design with four MPPs can support a precoding symbol throughput of 6.78 gigabits per second (Gbps) at a maximum frequency of 424 MHz,which is higher than that reported for K-best and FSE in \cite{barren'13}. It must also be noted that since ViPer NL-COMM can work much more reliably in lower SNRs than FSE, this means that the ViPer NL-COMM implementation can achieve much higher data throughput than FSE in those scenarios. Therefore, to evaluate the performance across varying real-world conditions, the average communication throughput over different SNRs provides a more comprehensive measure of efficiency, capturing system performance across a range of channel conditions. Mathematically, the average effective throughput ($\tilde{T}$) can be defined as
\begin{equation}
    \tilde{T}=\dfrac{1}{N}\sum_{i=1}^N T(1-PER(SNR_i))
\end{equation}
where $PER(SNR_i)$ is the packet error rate at SNR $i$, with the SNR ranging in 10-19 dB. The 4 $\times$ 4 ViPer NL-COMM accelerator improves the average effective throughput of VP K-best and FSE \cite{barren'13} by a factor of 1.65. More importantly, the 8 $\times$ 8 ViPer NL-COMM based NL precoder achieves an effective average throughput of 10.5 Gbps, demonstrating a substantial gain over existing solutions.

\begin{figure}[h]
    \centering
    \includegraphics[width=0.85\linewidth]{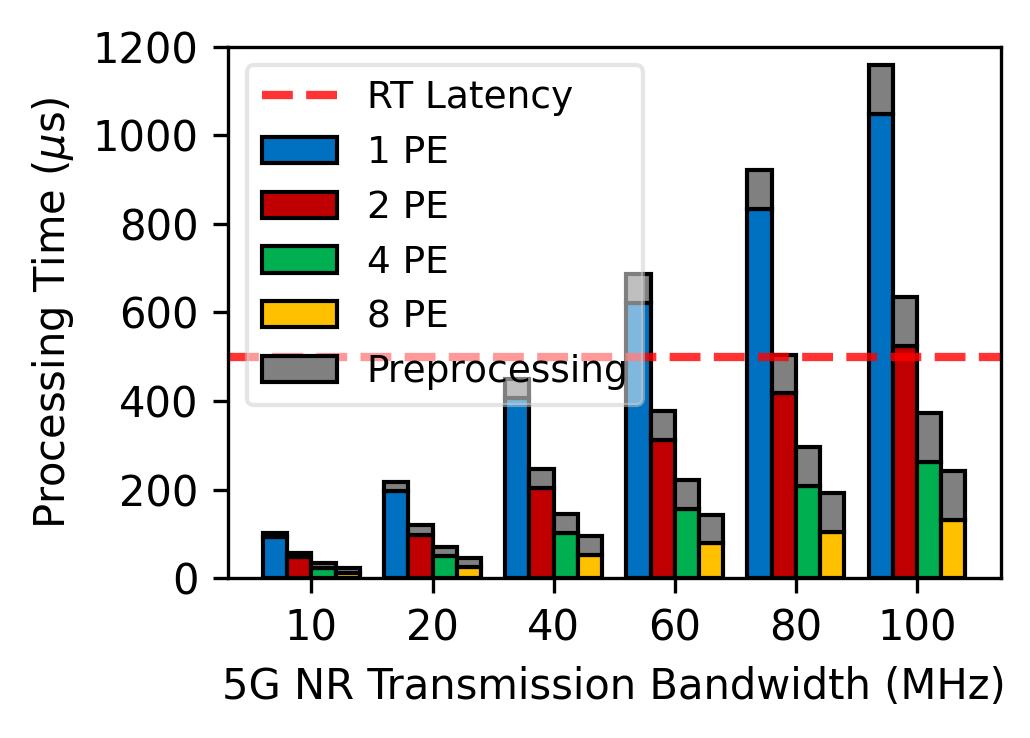}
    \caption{Indicative results of the processing latency for the FPGA implementation of the ViPer NL-COMM precoder with 8 MPPs for different PEs and 5G NR transmission bandwidths assuming 30 kHz subcarrier spacing.}
    \label{fig:latency_bw}
\end{figure}

\begin{figure}[h]
    \centering
    \includegraphics[width=0.85\linewidth]{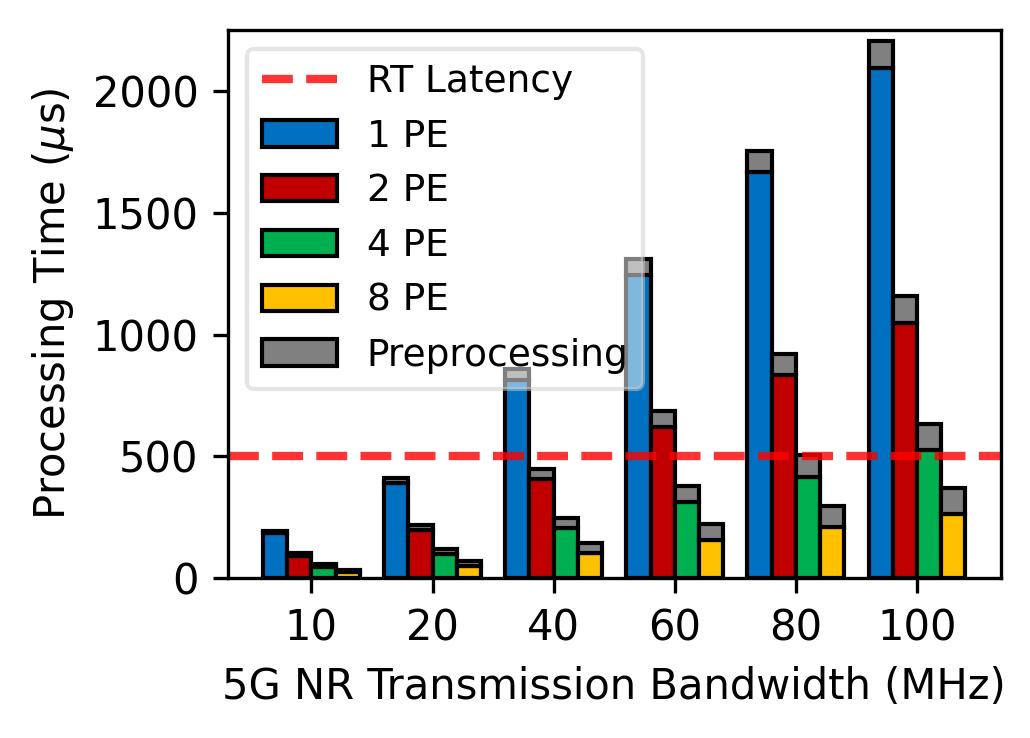}
    \caption{Indicative results of the processing latency for the FPGA implementation of the ViPer NL-COMM precoder with 16 MPPs for different PEs and 5G NR transmission bandwidths assuming 30 kHz subcarrier spacing.}
    \label{fig:latency_bw2}
\end{figure}

Fig. \ref{fig:latency_bw} shows the overall processing latency of the 8$\times$8 ViPer NL-COMM accelerator for different supported bandwidths and the number of instantiated PEs, shown relative to the 500, $\mu$s real-time (RT) slot budget (i.e., slot duration at 30 KHz subcarrier-spacing) as a baseline. Since the preprocessing is assumed to be performed once per slot, its associated latency is fixed for a specific bandwidth and becomes comparable to the postprocessing latency as the number of PEs increases. It can be observed that the accelerator can easily meet the latency constraints of current 5G-NR standards. Furthermore, the accelerator can be configured with different PEs for any desired bandwidth and latency. Fig. \ref{fig:latency_bw2} shows the required latency of the proposed 8$\times$ 8 accelerator to process 16 MPPs, which is achieved by multiplexing the paths up to a maximum of eight PEs, owing to the FPGA resource constraints. Even with an increase in processing time relative to Fig. \ref{fig:latency_bw}, the ViPer NL-COMM accelerator can still meet real-time constraints by scaling the number of parallel PEs for increasing bandwidths. 

\begin{figure}[h]
    \centering
    \begin{subfigure}{.1\textwidth}
    \centering
        \includegraphics[scale=.5]{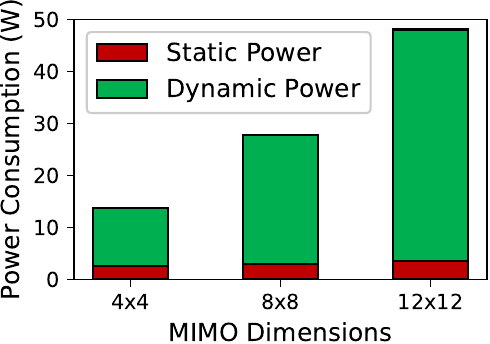}
        \caption{}
        \label{fig:pow}
    \end{subfigure}%
~
    \begin{subfigure}{.5\textwidth}
    \centering
        \includegraphics[scale=.5]{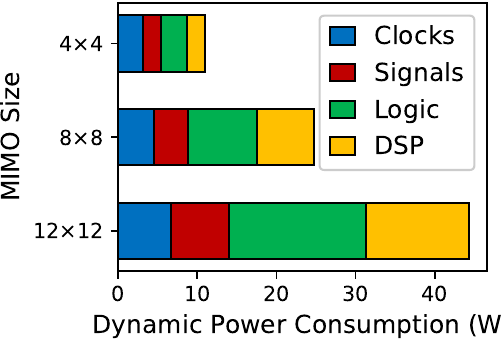}
        \caption{}
        \label{fig:pow2}
    \end{subfigure}
    \caption{(a) Static and dynamic power consumption of the FPGA for different MIMO sizes (b) Dynamic power consumption breakdown.}
    \label{fig:pow_overall}
\end{figure}

To evaluate the FPGA processing power required by the ViPer NL-COMM accelerator, Xilinx Power Estimator is used to analyze the design for different MIMO sizes. Fig. \ref{fig:pow} shows the power consumption of the corresponding ViPer NL-COMM accelerators with a split-up of the static and dynamic components of the power consumed. Fig. \ref{fig:pow2} shows how the different components of the dynamic power consumption, i.e., clocks, signals, logic, and DSP, scale with the MIMO size.

\subsection{PCIe Interfacing Overheads}\label{pcie_overhead}
Key interfacing concern relates to the transfer time of data between host and FPGA (ingress) and FPGA back to host (egress). For an 8×8 MU-MIMO system with 30 kHz SCS (500 µs/slot, 14 symbols/slot with 12 data symbols) and 32-bit complex samples, the required ingress payload data rate (i.e., for sending modulated I/Q for all data REs, and the channel estimates) is 0.783 GB/s at 20 MHz of transmission bandwidth and rises to  4.193 GB/s at 100 MHz. That is in the case where channel estimates are provided per subcarrier (i.e., one 8×8 matrix per subcarrier per slot). It is common in literature to relax the compute and transfer requirements by calculating the CSI on a per RB scale or larger \cite{hydra}. Under this softer assumption, the ingress rate drops to 0.496 GB/s (20 MHz) and  2.656 GB/s (100 MHz), respectively. Similarly, the egress rate (i.e., precoded vectors for the data REs) is 0.470 GB/s (20 MHz) and 2.516 GB/s (100 MHz). 
In terms of latency, the overhead of the corresponding serialized per-slot transfer times is: ingress 19.7 µs and egress 18.6 µs at 20 MHz, and ingress  105.4 µs and egress  99.8 µs at 100 MHz. That is under the assumption of a Gen3×16 link at ~80\% of line-rate (approximately 12.6 GB/s per direction).

While this overhead can be significant, in the majority of examined scenarios it remains well within our tight 500 µs slot budget (including the processing). Still, further reductions are feasible. First, offloading the modulation process to the FPGA, eliminates the need to transfer complex data vectors. This reduces ingress transfer latency per slot by 88.8\% (QPSK), 82.9\% (16-QAM), and 77.0\% (64-QAM) (bandwidth-independent). Second, as we discussed previously, by reducing the granularity of the channel estimates on a per-RB level, at the potential cost of performance loss, that is, if the channel varies significantly within an RB. The appropriate granularity should therefore be selected based on channel selectivity and the desired performance–I/O trade-off.

\subsection{Discussion}
\begin{figure}[h]
    \centering
    \includegraphics[width=0.9\linewidth]{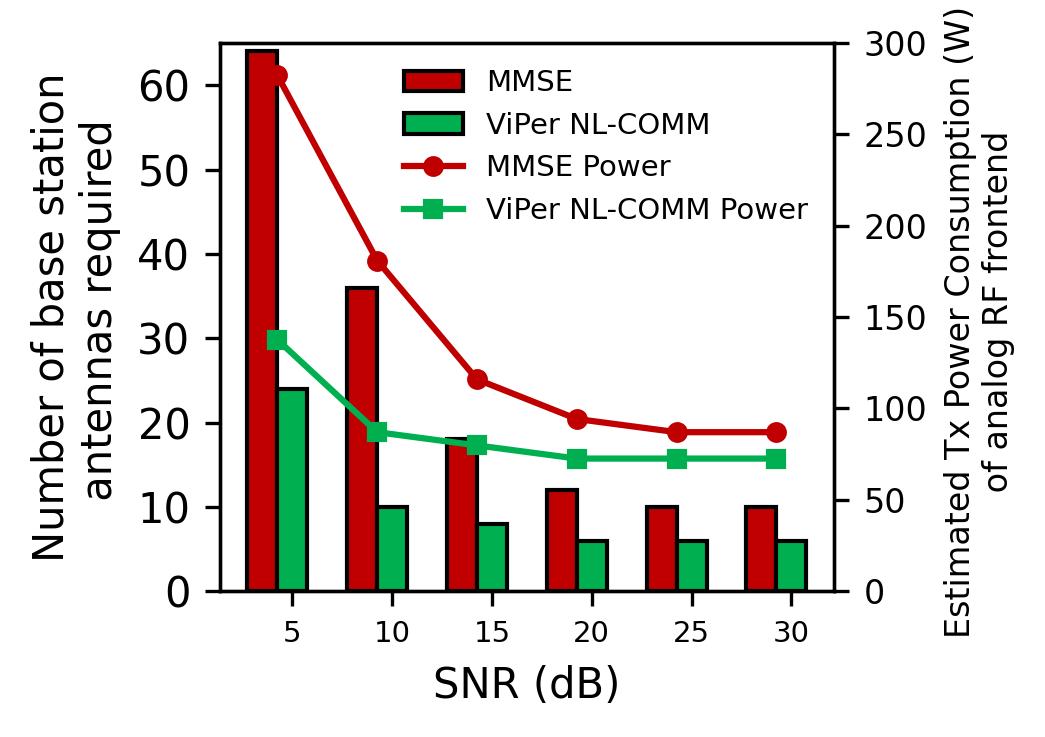}
    \caption{Comparison of the number of base station antennas required by MMSE and ViPer NL-COMM for different SNRs to meet the 10\% PER criterion for high-rate users (4 bps/Hz).}
    \label{fig:bs_snr64qam}
\end{figure}

In this subsection, we discuss the system-level gains that ViPer NL-COMM can provide in terms of the reduction in the number of base station antennas compared to traditional linear MMSE precoding. For this, we simulate different MIMO configurations over independent and identically distributed (i.i.d) Rayleigh channels. Firstly, we evaluate the performance of ViPer NL-COMM in supporting 8 high-rate users (64-QAM, 2/3 LDPC code rate, 4 bps/Hz) at different SNRs. Fig. \ref{fig:bs_snr64qam} shows that ViPer NL-COMM achieves AP antenna reduction factors of up to 2.67$\times$ at a lower SNR, while it saturates at 1.67$\times$ for a higher SNR. Moreover, ViPer NL-COMM enables marginal overloading of high-rate users at higher SNRs, supporting 8 such users with just 6 AP antennas, whereas MMSE does not enable any overloading.

\begin{figure}[h]
    \centering
    \includegraphics[width=.9\linewidth]{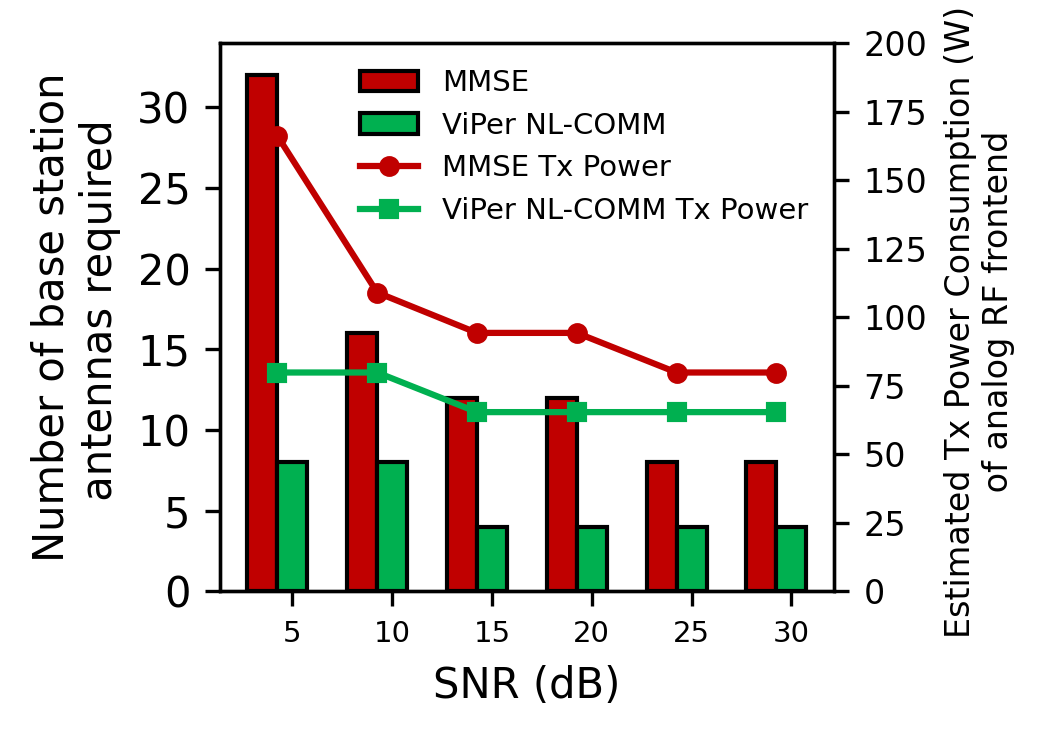}
    \caption{Comparison of the number of base station antennas required by MMSE and ViPer NL-COMM for different SNRs to meet the 10\% PER criterion for 8 concurrent low-rate users (1 bps/Hz).}
    \label{fig:bs_snr4qam}
\end{figure}

Next, we determine the minimum number of AP antennas required to support a fixed number of low-rate users (4-QAM, 1/2 LDPC code rate, 1 bps/Hz), achieving a packet error rate (PER) of 10\% or less for each user at different SNRs. Fig. \ref{fig:bs_snr4qam} shows that ViPer NL-COMM can substantially reduce the number of AP antennas in such scenarios and enable considerable overloading gains up to a factor of 2 for SNR within the range 15-30 dB, supporting 8 low-rate users with just 4 AP antennas. In the case of MMSE, even at a higher SNRs is not capable of delivering any overloading gains.

To translate the aforementioned reductions in the utilized antenna elements of ViPer NL-COMM into actual power consumption gains from the radio units, we utilize the model provided in \cite{halbauer'18} that estimates the power consumption values of the
different analog RF front-end building blocks.
Typically, the analog front-end, composed of transmit (Tx)/ receive (Rx) signal conversion, amplification, and filtering operations, is a major contributor to the overall power consumption budget of the MIMO system. Considering the typical values for power amplifier (PA) efficiency, total transmit power, insertion losses, and output power back-off presented in \cite{halbauer'18} for a realistic evaluation of different antenna array architectures summarized in Table \ref{tab:antenna_arrays}, we specifically assess the contribution of the downlink Tx PA and Tx signal conversion in the power consumption of a MIMO base station, assuming 256-QAM modulation, 3/4 LDPC code rate and 10 dB SNR. 

\begin{table}[h]
    \centering
    \caption{Examined Antenna Array Types \cite{halbauer'18}}
    \renewcommand{\arraystretch}{1.25}
    \resizebox{.95\linewidth}{!}{\begin{tabular}{|c|c|c|c|c|}
        \hline
        \textbf{Array} & \textbf{Total} & \textbf{Antenna} & \textbf{Subarray} & \textbf{Cross-Polarized } \\ 
        
        \textbf{ Type}&\textbf{Elements (N)} & \textbf{Ports (P)} & \textbf{Size (M\textsubscript{s})} & \textbf{Elements/Subarray}\\
        \hline
        
        A & 256 & 64 & 4 & 8 \\ 
        E & 128 & 64 & 2 & 4 \\ 
        F & 64 & 64 & 1 & 2 \\ 
        L & 64 & 16 & 4 & 4 \\ 
        \hline
    \end{tabular}}
    \label{tab:antenna_arrays}
\end{table}

\begin{figure}[h]
    \centering \includegraphics[scale=.45]{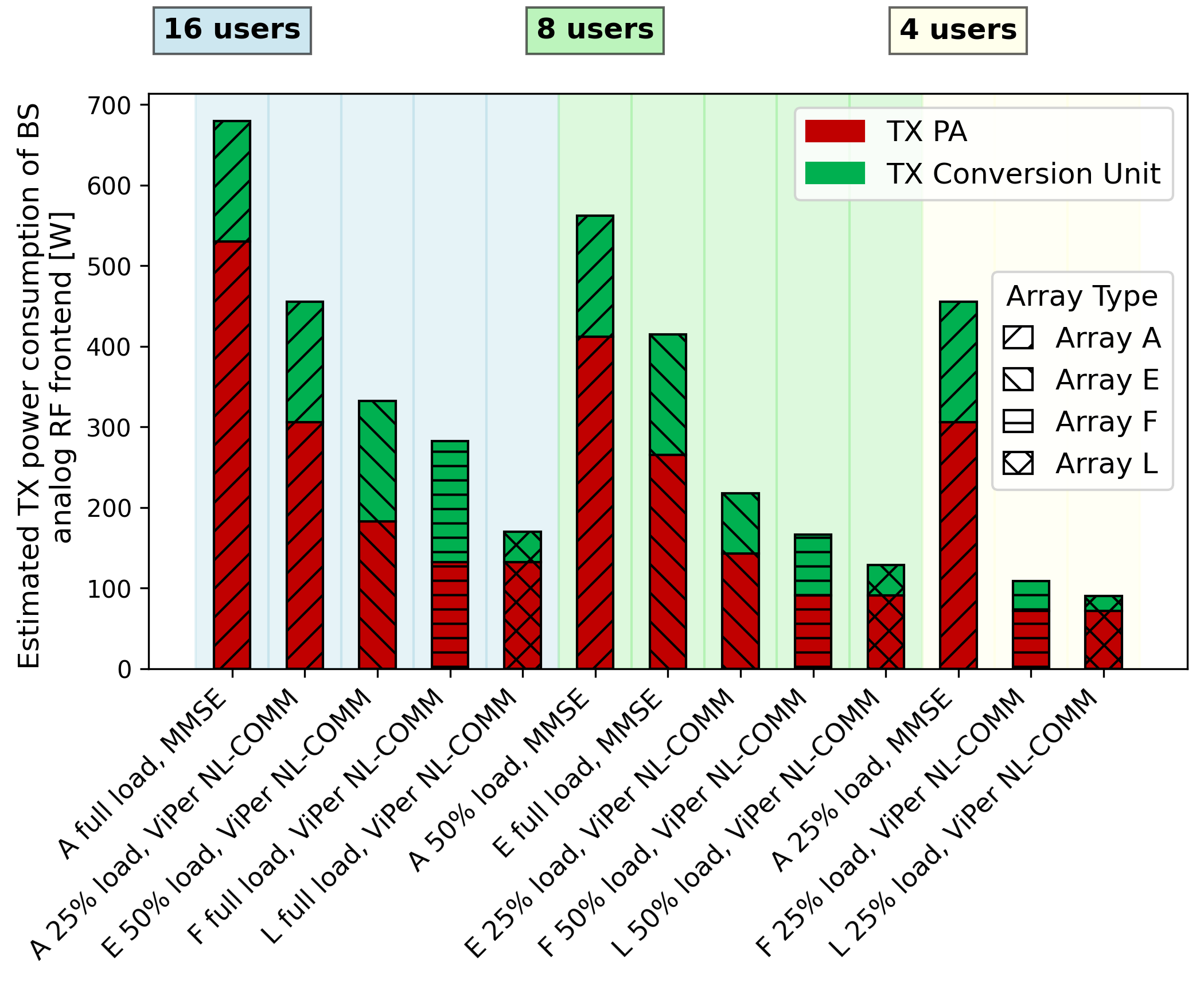}
    \caption{Power consumption of different array architectures for different number of supported users with MMSE and ViPer NL-COMM assuming a system with 100 MHz bandwidth and 53 dBm total transmit power for 256-QAM, LDPC code rate 3/4, 10 dB SNR. Load refers to the percentage utilization of the available antenna elements within the employed array.}
    \label{fig:bspow_comp}
\end{figure}

Fig. \ref{fig:bspow_comp} shows that ViPer NL-COMM can significantly reduce power consumption by employing different array types for the same number of users. In particular, MMSE precoding can support 16 users only with array type A with 256 antenna elements, while ViPer NL-COMM uses only 25\% elements of array A to support 16 users, which reduces the power consumption by 33\%. Using array types E, F and L with 50\%, 100\% and 100\% load respectively, ViPer NL-COMM can further reduce the power consumption by 27\%, 38\% and 62\%. It should be noted that MMSE precoding cannot support 16 users with any of these array types. For 8 users, ViPer NL-COMM can achieve 47\% power reduction over MMSE using array E. While MMSE cannot support F and L with full load conditions, ViPer NL-COMM achieves further power reductions by supporting F and L arrays with 50\% load. When 4 users are present in the system, ViPer NL-COMM can improve over MMSE's power consumption in arrays F and L by 61\% and 47\% respectively. 

\begin{figure}[h]
    \centering
    \includegraphics[width=0.9\linewidth]{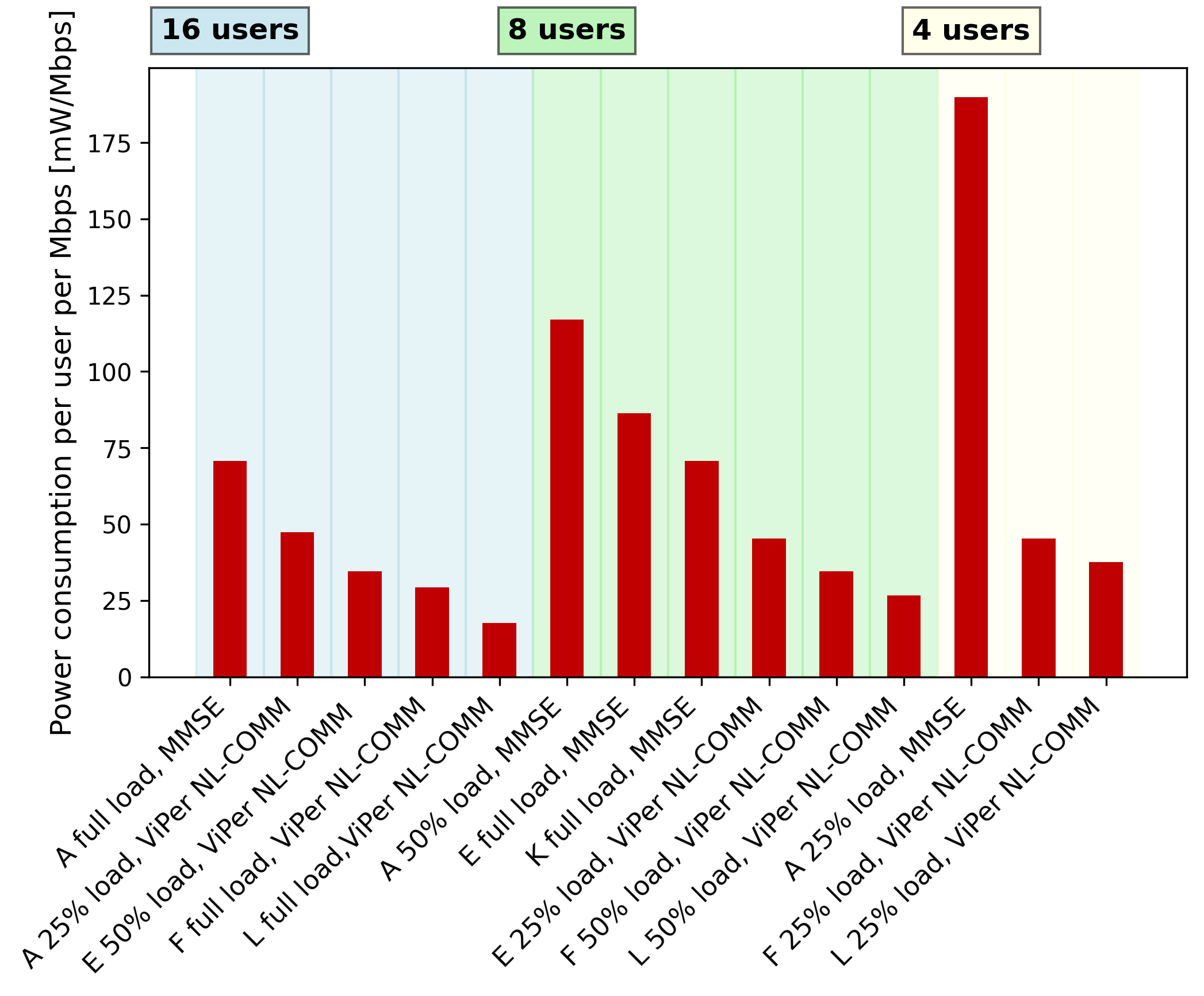}
    \caption{Power consumption per user per Mbps for different array architectures and loads.}
    \label{fig:bspoweff}
\end{figure}

Fig. \ref{fig:bspoweff} shows the evaluation of the power efficiency per user for different combinations of precoding algorithms and antenna array loads. It can be seen that ViPer NL-COMM achieves the highest power efficiency of around 17.6 mW per user per Mbps at 100\% load for the array L. It can also be seen that as the percentage of antenna elements activated is reduced, the cost in terms of power consumed per user per Mbps is increased compared to the fully activated case. This increase is more dramatic for array type A than E, F or L. 
\begin{figure}[h]
    \centering
    \includegraphics[width=0.9\linewidth]{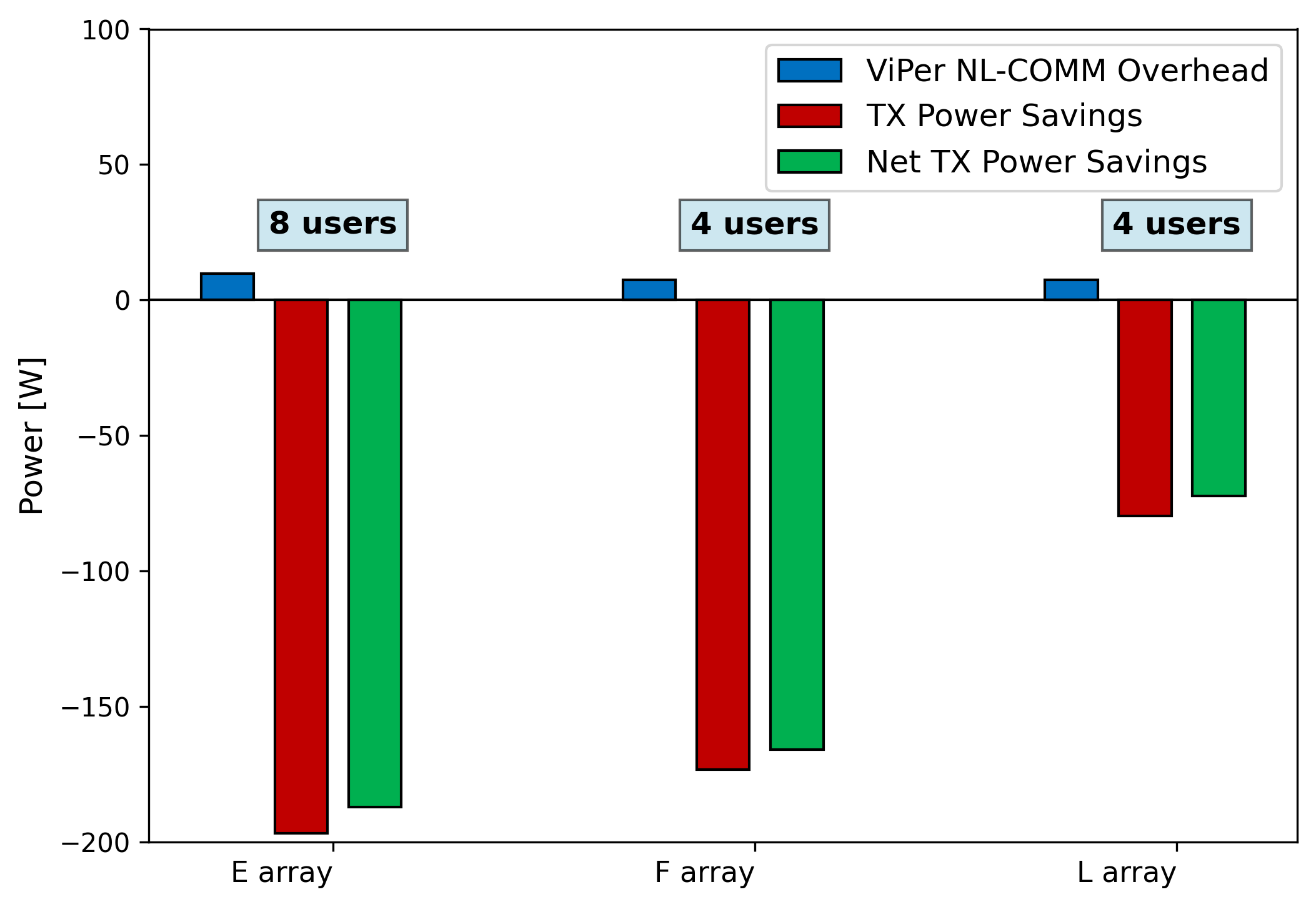}
    \caption{Net transmit (Tx) power savings of ViPer NL-COMM over MMSE for similar antenna array types and number of users.}
    \label{fig:fpgabspow}
\end{figure}
From a system perspective, we also analyze the FPGA power consumption overhead to ensure a fair comparison with linear ZF or MMSE precoding, excluding operations in ViPer NL-COMM that are analogous to those in MMSE, such as QR decomposition and precoding matrix multiplication. As shown in Fig. \ref{fig:fpgabspow}, ViPer NL-COMM enables much lower loading factors than MMSE, resulting in significant power savings. For instance, ViPer NL-COMM can support 8 users with a 25\% loaded array E, whereas MMSE requires E to be fully loaded to ensure 8 users are supported. The FPGA overhead for ViPer NL-COMM processing is much smaller than the induced savings, leading to 47\%, 61\%, and 47\% reduction in the Tx power consumption of the analog RF frontend.

\section{Conclusion \& Future Directions}
This work presents, for the first time, ViPer NL-COMM, a non-linear precoding approach which fully exploits the available channel capacity and can support even more users than the number of AP antennas. The proposed approach achieves higher throughput and lower latency at much lower error rates than all existing tree-search precoders. The first ViPer NL-COMM FPGA implementation can support in real-time wide transmission bandwidth up to 100 MHz and high MIMO dimensions of up to 16$\times$16. These initial results demonstrate the practical applicability of advanced physical layer processing in real systems where spectral and energy efficiency are critical. This is particularly relevant for emerging Open RAN deployments, where the inherent architectural flexibility can, as discussed in this work, further amplify the gains offered by ViPer-NLCOMM.

Future developments will aim to leverage reconfiguration of the RF channel \cite{hoffman'24} to achieve dynamic fine-grained activation and deactivation of antenna elements and RF chains in tune with system requirements. Moreover, the flexible processing enabled by ViPer NL-COMM in terms of PEs employed can be adapted to the actual environment, optimizing computational power consumption to meet the required error rates.
In Open RAN deployments, adaptive control can be implemented within RAN Intelligent Controller (RIC) applications (i.e., xAPPs or rAPPs). These applications use real-time RAN statistics, such as live power measurements from O-RUs/O-DUs, to make intelligent RF reconfiguration decisions. For instance, they can optimize computational and RF power consumption by dynamically increasing the number of active antennas on the O-RU or allocating additional PEs to the baseband processing unit as traffic demand grows. This flexibility allows for real-time adaptation to power budgets, QoS demands, and network conditions, enabling more sustainable and efficient Open RAN operations.

\bibliographystyle{IEEEtran}
\bibliography{mybib}

\break

\begin{IEEEbiographynophoto}{Thomas James Thomas}
received the Ph.D. degree in VLSI signal processing from Indian Institute of Space Science and Technology, Thiruvananthapuram. He has been a Research Fellow with the 6G Innovation Center (6GIC), University of Surrey, U.K. His research interests include the design of advanced signal processing architectures for real-time applications focused on future wireless and open RAN
communication systems.
\end{IEEEbiographynophoto}
\vskip -2\baselineskip plus -1fil

\begin{IEEEbiographynophoto}{George N. Katsaros} (Member, IEEE)
is a Research Fellow at the 6G Innovation Center (6GIC), University of Surrey, UK, where he also received his Ph.D. He holds a B.Sc. and M.Sc. in Physics from the University of Athens, Greece. His research focuses on the architectural and algorithmic design of real-time, power-efficient physical layer (PHY) processing.
\end{IEEEbiographynophoto}
\vskip -2\baselineskip plus -1fil

\begin{IEEEbiographynophoto}{Chathura Jayawardena}(Member, IEEE)
received the Ph.D. degree from the University of Surrey, Guildford, U.K. He is currently a Research Fellow with the University of Surrey. His research interest includes signal processing for communications, with an emphasis on detection methods for non-orthogonal transmission schemes.
\end{IEEEbiographynophoto}
\vskip -2\baselineskip plus -1fil

\begin{IEEEbiographynophoto}{K. Nikitopoulos}(Senior Member, IEEE)
is currently a Professor with the
Institute for Communication Systems, University
of Surrey, U.K., and the Director of its “Wireless Systems Lab.” He is an Active Academic Member of the 6G Innovation Centre (6GIC), where he leads the “Theory and Practice of Advanced Concepts in Wireless Communications” research area. His research focuses on advanced signal processing, next-generation computing architectures, and system-level wireless design. He is the lead inventor of NL-COMM (www.nl-comm.com). He is a Senior Member of IEEE and recipient of the prestigious First Grant of the U.K.’s Engineering and Physical Sciences Research Council.

\end{IEEEbiographynophoto}
\vskip -2\baselineskip plus -1fil

\end{document}